\documentclass[a4paper, 11pt]{article}
\usepackage{my_jcap} 

\usepackage{natbib}
\setcitestyle{aysep={}}

\usepackage{graphicx}	
\usepackage{amsmath}	

\usepackage{fontawesome5}
\usepackage{color}
\usepackage{xcolor}

\usepackage{lineno}

\newcommand{\msol}{\ensuremath{\, {\rm M}_\odot}}

\newcommand{\mpc}{\ensuremath{\, {\rm Mpc}}}         
\newcommand{\gpc}{\ensuremath{\, {\rm Gpc}}}

\newcommand{\eg}{{\sl e.g.}, }

\newcommand{\CDF}{{\rm CDF}}

\newcommand{\nhat}{\mathbf{\hat{n}}}

\newcommand*{\vcenteredhbox}[1]{\begingroup
\setbox0=\hbox{#1}\parbox{\wd0}{\box0}\endgroup}

\newcommand{\OrcidID}[1]{ \href[urlcolor = red]{https://orcid.org/#1}{\textcolor{lightgray}{\faOrcid}}}
\newcommand{\OrcidIDName}[2]{\href{https://orcid.org/#1}{#2}}

\newcommand{\fNLLoc}{f_{\rm NL}^{\rm \,loc}}
\newcommand{\fNLOR}{f_{\rm NL}^{\rm \,or,\, lss}}
\newcommand{\fNLEQ}{f_{\rm NL}^{\rm \,eq}}
\newcommand{\fNLORCMB}{f_{\rm NL}^{\rm \,or,\, cmb}}
\newcommand{\fNL}{f_{\rm NL}}

\newcommand{\kOne}{k_1}
\newcommand{\kTwo}{k_2}
\newcommand{\kThree}{k_3}

\newcommand{\Loc}{{\rm loc}}
\newcommand{\OR}{{\rm or}}
\newcommand{\EQ}{{\rm eq}}
\newcommand{\ORCMB}{{\rm or, cmb}}

\title{Primordial non-Gaussianities with weak lensing: Information on non-linear scales in the \textsc{Ulagam} full-sky simulations}

\author[1, 2]{\OrcidIDName{0000-0003-3312-909X}{Dhayaa Anbajagane}
(\vcenteredhbox{\includegraphics[height=1.2\fontcharht\font`\B]{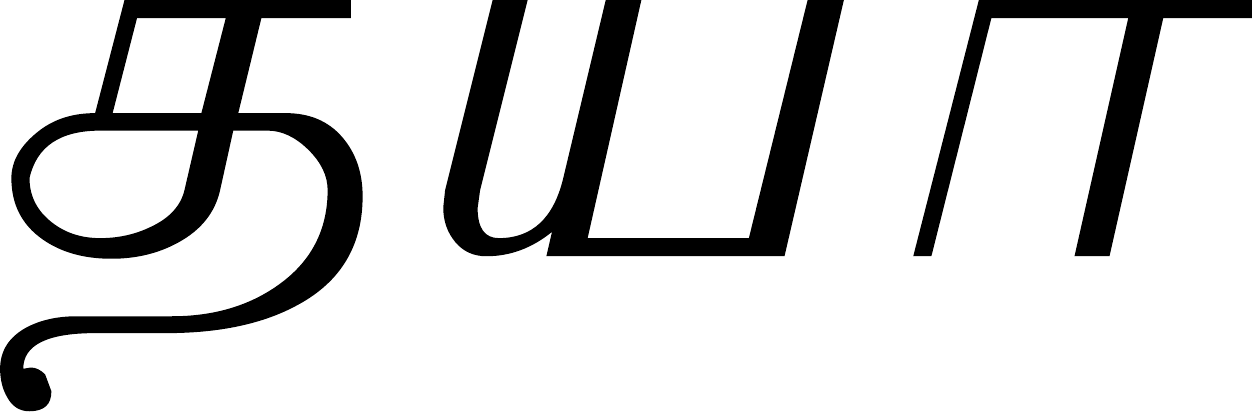}}),}
\author[1,2]{\OrcidIDName{0000-0002-7887-0896}{Chihway Chang},}
\author[2]{Hayden Lee,}
\author[3]{Marco Gatti}

\affiliation[1]{Department of Astronomy and Astrophysics, University of Chicago, Chicago, IL 60637, USA}
\affiliation[2]{Kavli Institute for Cosmological Physics, University of Chicago, Chicago, IL 60637, USA}
\affiliation[3]{Department of Physics and Astronomy, Center for Particle Cosmology, University of Pennsylvania, Philadelphia, PA 19104, USA}

\emailAdd{dhayaa@uchicago.edu}

\abstract{
Primordial non-Gaussianities (PNGs) are signatures in the density field that encode particle physics processes from the inflationary epoch. Such signatures have been extensively studied using the Cosmic Microwave Background, through constraining their amplitudes, $f^{X}_{\rm NL}$, with future improvements expected from large-scale structure surveys; specifically, the galaxy correlation functions. We show that weak lensing fields can be used to achieve competitive and complementary constraints. This is shown via the \textsc{Ulagam} suite of N-body simulations, a subset of which evolves primordial fields with four types of PNGs. We create full-sky lensing maps and estimate the Fisher information from three summary statistics measured on the maps: the moments, the cumulative distribution function, and the 3-point correlation function. We find that the year 10 sample from the Rubin Observatory Legacy Survey of Space and Time (LSST) can constrain PNGs to $\sigma(\fNLEQ) \approx 110$, $\sigma(\fNLOR) \approx 120$, $\sigma(\fNLLoc) \approx 40$. For the former two, this is better than or comparable to expected galaxy clustering-based constraints from the Dark Energy Spectroscopic Instrument (DESI). The PNG information in lensing fields is on non-linear scales and at low redshifts ($z \lesssim 1.25$), with a clear origin in the evolution history of massive halos. The constraining power degrades by $\sim\!\!60\%$ under scale cuts of $\gtrsim 20\mpc$, showing there is still significant information on scales mostly insensitive to small-scale systematic effects (\eg baryons). We publicly release the \textsc{Ulagam} suite to enable more survey-focused analyses.}

\begin{document}
\maketitle
\flushbottom



\section{Introduction}

An overarching goal of cosmology is to understand the history of the Universe, both its initial state and its subsequent evolution to the present epoch. The current paradigm for the generation of the initial density field (i.e. the initial state) is inflation, a mechanism that generates quantum fluctuations during an epoch of rapid exponential expansion \citep[see][for a review]{Guth2004Inflation}. This density field is then evolved under a $\Lambda \rm CDM$ model, where CDM stands for cold dark matter and $\Lambda$ is the cosmological constant. The large-scale structure --- which is the distribution of matter in the Universe --- depends on both the initial conditions and their subsequent evolution, and is thus a useful probe for studying the full history of the Universe. Analyses of this structure, as traced at high redshift by the cosmic microwave background (CMB) or at low redshift by galaxy surveys, have already shed light on the properties of our Universe and on the values of the six parameters that make up the $\Lambda \rm CDM$ model \citep{Planck2016CosmoParams,  Asgari2021Kids1000, DES2022Y3, More2023HSCY3}. Other analyses have probed the physics of the initial conditions, and in particular, have led to constraints disfavoring certain classes of inflationary models \citep{Planck2020Inflation}.

In the simplest models, only a single field --- commonly called the ``inflaton'' field --- is present during inflation and slowly rolls down the potential, resulting in simple, weak interactions. In this case, the initial density field generated is highly Gaussian \citep{Maldacena2003SingleFieldPNGs}. Such a field is defined entirely by the covariance of densities in different parts of the field, which is a Power spectrum in Fourier space or a 2-point correlation function in real space. Such functions capture the correlations between any two points in a field (or two different fields) separated by a given distance. By adding complexity to the inflation model --- such as additional fields that interact with one another, higher-order interactions within a single field, and so forth --- the inflationary mechanisms gain non-linear terms that then lead to primordial non-Gaussianities (PNG) in the initial density field. The amplitudes of the PNGs are captured by the $\fNL$ parameters, and directly probe various energy scales in the theory of inflation~\citep[][see their Figure 1]{Achucarro2022InflationReview}. Given that inflation might have taken place at very high energies of $10^{14}\,\, {\rm GeV}$ (giga electron volts), which is much larger than energy scales achievable in terrestrial particle physics experiments, the parameter $\fNL$ probes what has been denoted an ``energy frontier'' in cosmology \citep{Achucarro2022InflationReview}. Note that $\fNL$ only captures the leading deviation from Gaussianity and corresponds to a bispectrum or three-point correlation, which defines the correlation between \textit{three} points in the field (or three different fields) represented in either Fourier space or real space, respectively. One can view such correlations as generating a skewness in the field, though this is not a formal equivalence and simply serves as a useful heuristic. Higher-order correlations, such as the 4-point function parameterized by $\tau_{\rm NL}$,
can also be non-zero due to inflationary mechanisms. However, they are not the focus of this work.

The current best constraints on $\fNL$ are found in the bispectrum analysis of the \textit{Planck} CMB data \citep{Planck2020PNGs} --- $\sigma(\fNLEQ) = 5$, $\sigma(\fNLEQ) = 47$, $\sigma(\fNLORCMB) = 24$ --- as the spatial correlations of the observed temperature fluctuations arise from inflationary correlations. These constraints are primarily limited by cosmic variance, rather than by measurement noise. Future CMB surveys such as CMB S4 \citep{CMBS42019} can only modestly improve on this result through reduction of the measurement noise; the \textit{Planck} data already covers most of the observable sky, thus CMB-S4 will not improve on the cosmic variance limit. However, more significant improvements are expected from large-scale structure (LSS) galaxy surveys, using the correlations of galaxy positions as a probe of the inflationary correlations. These surveys probe a 3D volume, resulting in the number of countable modes scaling as $N_k \propto Vk_{\rm max}^3$, while in a 2D CMB map, the number of modes scales as $N_k \propto Ak_{\rm max}^2$. Here, $V$ is the 3D survey volume, $A$ is the 2D map area, and $k_{\rm max}$ is the highest $k$-mode studied in the analysis. As galaxy surveys push to higher redshift (which increases the survey volume) and higher resolution (which improves the smallest measured scale), the number of measured modes increases significantly. Thus, galaxy survey measurements will have superior statistical power to CMB measurements, and can then be used to constrain inflationary physics. In addition, galaxy correlations probe a noticeably different set of length scales than the CMB, and the combination of the two can probe scale-dependent $\fNL$ models (see Section \ref{sec:simlensing}). Many works have extracted PNG constraints from galaxy correlation measurements \citep{Mueller2021BOSSDr16fNL, Cabass2022SingleFieldBOSS, Cabass2022MultifieldBOSS, Damico2022BossPNG, Philcox2022BossPNG}.

Another cosmological probe observed by many of the same widefield galaxy surveys is weak lensing, which is distortions in the shapes of galaxies --- commonly denoted ``source galaxies'' --- due to the foreground structure present between the observer and the galaxies \citep{Bartelmann2001WLReview}. The spatial correlations of these distortions are generated by the foreground density field and are thus a probe of cosmology. While all observational constraints on $\fNL$ from galaxy surveys have focused on the 3D galaxy field, none have focused on the weak lensing field (though there exists some theoretical work in this direction as we discuss below). There are a number of complementary benefits in using weak lensing as a probe of inflation, such as the lensing field being a direct, unbiased tracer of the density field insensitive to the physics of the galaxy--halo connection,\footnote{As we will discuss later in Section \ref{sec:ModelChallenge}, weak lensing is still sensitive to baryonic physics given the latter impacts the density field that generates the weak lensing signal. However, this is a distinctly different kind of phenomenon from the galaxy--halo connection, which concerns itself with the distribution of all galaxies that can populate a given halo, and thus involves smaller-scale physics.} efficient simulation-based modeling of strongly non-linear scales enabled by less stringent resolution requirements, and so on (a more detailed discussion is presented in Section \ref{sec:simlensing}). These advantages are currently not being utilized in studies of inflationary physics as lensing-based analyses are yet to be implemented. 

Historically, a limitation in performing such lensing-based studies has been obtaining a computationally efficient model for $\fNL$ signatures in weak lensing. Given that the weak lensing signal probes a line-of-sight integral of the density field, a majority of the measurements contain some contributions from the non-linear density field \citep[\eg][see their Figure 4]{Secco2022Shear}.
Galaxy correlations from $\fNL$ signals have been efficiently modelled using the effective field theory of large-scale structure (EFTofLSS), which is an analytic approach to modeling the correlations of the density field and the galaxy field \citep{Baumann2012EFTofLSS, CarrascoEFTofLSS}. The current calculations of the two-loop power spectrum and one-loop bispectrum are accurate up to quasi-linear scales and have significant deviations ($\geq 10\%$) in the non-linear regimes ($k \gtrsim 1 h/{\rm Mpc}$); for example, see \citet[][their Figure 10]{Baldauf2015EFTvsNbody} or \citet[][their Figures 2-7]{Sefusatti2010EFTvsNbody}. Therefore, it is difficult to employ the EFT approach to model weak lensing. However, over the past decade, the feasibility of full, simulation-based modeling has grown significantly and has led to multiple analyses of the lensing field that are simulation-based \citep[\eg][]{Fluri2018DeepLearning, Fluri2019DeepLearningKIDS, Zurcher2021WLForecast, Fluri2022wCDMKIDS,Zurcher2022WLPeaks}, or more often at least simulation-informed \citep[\eg][]{Secco2022Shear, Amon2022Shear, Gatti2022MomentsDESY3}. Thus, with modern advancements in computing, it is now possible to efficiently and accurately model the non-linear density field through N-body simulations \citep[see][for a review]{Angulo2022SimReview}, which thereby enables analyses of $\fNL$ with weak lensing.

Previous works have theoretically explored the power of weak lensing in constraining PNGs, and have found it to be a potentially promising probe \citep{Marian2011fNL, Shirasaki2012fNL, Hilbert2012fNL}. These works used a modest number of simulations to estimate the signal of a specific type of $\fNL$, called the ``local'' type (see Section \ref{sec:Simulations}) on the lensing field, and used simple scaling arguments for how constraining power increases with survey area to approximately estimate the constraints from wide-field lensing surveys. However, as discussed above, it is now possible to run substantially larger suites of large-volume high-resolution simulations of the full sky and estimate the inflationary information in the lensing field. In addition, there are other types of $\fNL$ --- beyond the local type --- that have valuable information about inflation and have not been considered in simulation-based analyses of weak lensing, though some analytical approaches have been previously employed \citep{Schafer2012PNGwithWL,  Giannantonio2012fNLSurvey}.

In this work, we explore the use of the lensing convergence field as a probe of PNGs. We expand on previous efforts by (i) developing and using a large simulation suite ($N_{\rm sims} = 3600$), which enables better numerical accuracy of Fisher information estimates and allows a closer match of lensing survey specifications, (ii) exploring four types of $\fNL$, each of which probes different primordial physics and three of which are being studied for the first time in the context of simulation-based constraints from weak lensing on non-linear scales, and (iii) forecasting realistic constraints for current and upcoming widefield surveys using their expected redshift distributions, noise amplitudes, survey area etc. This work is organized as follows: in \S \ref{sec:Simulations} we describe the suite of simulations developed for this work, including the types of PNGs we focus on, and also the forward modeling procedure to simulate the weak lensing observations from each survey. In \S \ref{sec:HOS} we describe the statistics used to summarize the lensing field, which includes the moments, cumulative distribution functions, and the three-point function. We present our results in \S \ref{sec:Results}, including the Fisher information in different statistics and surveys, and the physical origin of the PNG signal in lensing. We discuss the advantages and caveats of lensing-based analyses of inflation in \S \ref{sec:Discussion}, and then conclude in \S \ref{sec:Conclusion}.

\section{Simulations}\label{sec:Simulations}

The PNGs, by virtue of their impact on the initial density field, can affect non-linear structure formation and imprint onto any field related to this structure, such as the galaxy number density fields as in the studies discussed above. In this work, we are interested in the lensing convergence field, $\kappa$, which is a line-of-sight integral of the density field,
\begin{equation}\label{eqn:convergence_definition}
    \kappa(\nhat, z_s) = \frac{3}{2}\frac{H_0^2\Omega_{\rm m}}{c^2}\int_0^{z_s}\!\!\!\delta(\nhat, z_j) \frac{\chi_j(\chi_s - \chi_j)}{a(z_j)\chi_s}dz_j\frac{d\chi}{dz}\bigg|_{z_j},
\end{equation}
where $z_s$ is the redshift of the ``source'' plane/galaxies being lensed, $\nhat$ is the pointing direction on the sky, $\delta$ is the overdensity field, $\chi$ is the comoving distance from an observer to a given redshift, $a$ is the scale factor, $H_0$ is the Hubble constant, $\Omega_{\rm m}$ is the matter energy density fraction at $z = 0$, and $c$ is the speed of light. We use the shorthand $\chi(z_s) \equiv \chi_s$ and $\chi(z_j) \equiv \chi_j$.

We model this convergence --- including its dependence on PNGs and cosmology --- using full-sky density maps from N-body simulations. Such simulations are uniquely suited for modeling these fields in the non-linear regime. The set of simulations introduced in this work will henceforth be referred to as the \textsc{Ulagam} suite.\footnote{\textit{Ulagam} is a Tamil word (pronounced ``\textit{ooh-luh-gum}'') that denotes the World or the Universe, and this naming choice is inspired by the \textsc{Uchuu} simulations --- a set of multi-${\rm Gpc}$, high-resolution simulations ---  which were named after the Japanese word for Universe.} The simulations are run with the \textsc{Pkdgrav3} solver \citep{Potter2017Pkdgrav3}, which has been used extensively in modeling the weak lensing field \citep{Fluri2018DeepLearning, Fluri2019DeepLearningKIDS, Zurcher2021WLForecast, Zurcher2022WLPeaks, Gatti2022MomentsDESY3, Kacprzak2023Cosmogrid, Gatti2023SC}. \textsc{Pkdgrav3} automatically builds lightcones --- where these cones are built using the methods first described in \citet{Fosalba2008Shells, Das2008LensingShells} --- while solving the gravitational dynamics of the system forward in time, and so our final outputs are the lightcone shells (i.e. \textsc{Healpix} maps) of the density field at different redshifts. The simulation box is tiled/repeated as needed to construct large enough volumes to then build full-sky lightcones to a given redshift. This repetition will bias any large-scale correlations in the lightcone, but in this work we only consider scales much smaller than the box size.

These simulations are run in $L = 1 \gpc/h \approx 1.5 \gpc$ boxes, starting at $z = 127$, and with $N = 512^3$ dark matter particles. The initial conditions for all runs are obtained from the \textsc{Quijote} suite \citep{Navarro2020Quijote} and the \textsc{Quijote-png} extension \citep{Coulton2022QuijotePNG}. Thus, these simulations are lightcone companions of the \textsc{Quijote} simulations and are specialized for widefield survey analyses. The original \textsc{Quijote} suite was designed for studying the Fisher information of non-linear structure, as well as for building emulators sampling different cosmological parameters, but the available data products provide inadequate redshift resolution for producing accurate mock lightcones of the lensing/density field. Hence we have resimulated a subset of these simulations to create accurate full-sky density and lensing maps. When running simulations with PNGs, these PNGs are included in the density field of the initial conditions --- using the techniques described below in Section \ref{sec:PNG_intro} --- and the field is then evolved via the fiducial N-body solver with no modifications.

The \textsc{Ulagam} suite contains simulations for computing the derivatives of the lensing/density field with respect to multiple $w{\rm CDM}$ and PNG parameters; these are $\Omega_{\rm m}$, $\sigma_8$, $w$, and $n_s$ for the $\Lambda {\rm CDM}$ case and different $\fNL$ corresponding to four types --- local, equilateral, orthogonal LSS and orthogonal CMB --- which are detailed in Section \ref{sec:PNG_intro}. The suite contains 100 simulations per parameter where the value of that parameter is slightly higher than the fiducial, and another 100 simulations where the value is slightly lower than the fiducial, and these paired sets are used to compute the derivatives. The fiducial cosmology is from \citet{Planck2016CosmoParams}, and the derivatives are computed over differences of $\Delta \Omega_{\rm m} = 0.02$, $\Delta \sigma_8 = 0.03$, $\Delta w = 0.05$, $\Delta n_{\rm s} = 0.04$, and $\Delta \fNL^{\rm type} = 200$, which are all the same settings as the \textsc{Quijote} and \textsc{Quijote-PNG} suites. The \textsc{Ulagam} suite also contains $2000$ simulations at the fiducial cosmology which can be used to compute the covariance matrix for data-vectors. Each all-sky map can have multiple, completely independent cutouts of the survey footprints, so in practice, the simulations provide between 6000 to 8000 realizations for the covariance, and also between 300 to 400 realizations for the derivatives; the lower and upper bound numbers are for LSST and DES respectively. A summary of the available full-sky runs is listed in Table \ref{tab:SimProducts}.

The simulations have a total of 100 timesteps/shells, with 95 shells between $0 < z < 10$, and a high redshift resolution of $\Delta z \approx 0.01 - 0.05$ in the latter redshift range; the exact value of $\Delta z$ depends on the shell. The timesteps in this redshift range are spaced uniformly in proper time, $t$, and this corresponds to different $z$ and comoving distances depending on the cosmology. The density shells are then post-processed via Equation \ref{eqn:convergence_definition}, with the integral over $z_j$ now replaced by a simple discrete sum, to create a lensing convergence field at each source plane redshift, $z_s$. This technique uses the Born approximation, which computes the total effective deflection due to lensing but along an undeflected ray path. A more precise calculation uses full raytracing, which calculates these deflections while also constantly deflecting/updating the ray path. \citet{Petri2017Born} found the Born approximation leads to differences of $\lesssim 5\%$ for the third moments statistic we will use in Section \ref{sec:Results}. This effect is important when requiring a certain \textit{absolute} accuracy in the simulation predictions, whereas for estimating the Fisher information --- as we do in this work --- these requirements can be relaxed given we only require accuracy in the \textit{relative} differences between different simulations (as discussed further below).

Halos are identified in the 3D simulation volume using a friends-of-friends (FoF) percolation technique with a linking length of $b = 0.2$, in units of the mean inter-particle distance, consistent with the choice made for \textsc{Quijote}. The halo catalogs contain all identified objects with at least 100 particles, which corresponds to a minimum mass of $M > 10^{14} \msol$ across all redshifts. This is different from the \textsc{Quijote} catalogs, which keep all halos down to 20 particles. The change was made to reduce the total storage footprint of the simulation suite.

While the original \textsc{Quijote} suite was run using \textsc{Gadget3} \citep[last described in][]{Springel2005Gagdet2}, we use \textsc{Pkdgrav3} in this work given its specialization and extensive use in modeling full-sky observables. We verify in Figure \ref{fig:QuijoteComparison} that we can reproduce the results from the \textsc{Quijote} simulations to within expected accuracy given the difference in the two N-body solvers. Similarly, we have also performed checks on the choices of particle count in Appendix \ref{appx:Validation}. The numerical requirements for this work are less stringent than a full simulation-based model as we do not use the simulations for cosmological inference, but rather for (i) computing derivatives for the Fisher information, where the relevant quantities are relative differences in the simulations as we vary cosmological parameters, and for (ii) computing covariance matrices for the Fisher information, where once again the relevant quantities are relative differences in simulations across different realizations. As a result, our requirements for absolute accuracy/calibration of the simulations are relaxed. Thus, while some results in Appendix \ref{appx:Validation} suggest the simulations' accuracy can benefit from a higher particle count, the current suite is still adequate for estimating the Fisher information --- which was the original purpose of the \textsc{Quijote} simulations that the \textsc{Ulagam} suite now builds on.

\begin{table*}
    \centering
    \begin{tabular}{l|c|c}
    \hline
      Run & $\mathbf{P_{\rm fid}} \pm \Delta P$ & $N_{\rm sim}$ \\[5pt]
      \hline
      Fiducial & \textbf{---} & \textbf{2000}\\[5pt]
      Local PNG, $\fNLLoc$ & $\mathbf{0} \pm 100$ & 100\\[5pt]
      Equilateral PNG, $\fNLEQ$ & $\mathbf{0} \pm 100$ & 100\\[5pt]
      LSS Orthogonal PNG, $\fNLOR$ & $\mathbf{0} \pm 100$ & 100\\[5pt]
      CMB Orthogonal PNG, $\fNLORCMB$ & $\mathbf{0} \pm 100$ & 100\\[5pt]
      Matter density, $\Omega_{\rm m}$ & $\mathbf{0.3175} \pm 0.01$ & 100\\[5pt]
      Density fluctuations amplitude. $\sigma_8$  & $\mathbf{0.834} \pm 0.015$ & 100\\[5pt]
      Dark energy EoS $w_0$ & $\mathbf{-1} \pm 0.05$ & 100\\[5pt]
      Spectral index $n_s$ & $\mathbf{0.9624} \pm 0.02$ & 100\\[5pt]
      \hline
    \end{tabular}
    \caption{The simulation runs presented in this work. The fiducial simulation parameters follow those of \citet{Planck2016CosmoParams} and are shown in bold, while the variations to the parameters for calculating derivatives are shown as the $\pm \Delta P$ values. The cosmological parameter values used in the fiducial runs are all the bolded values. We always assume a flat cosmology with $\Omega_\Lambda = 1 - \Omega_{\rm m}$. The parameters not shown above are $h = 0.6711$ and $\Omega_{\rm b} = 0.049$.}
    \label{tab:SimProducts}
\end{table*}

\subsection{Initial conditions with primordial non-Gaussianities }\label{sec:PNG_intro}

Generating initial conditions for a purely Gaussian initial density field is a well-studied procedure with established numerical recipes \citep[\eg][]{Crocce2006LPT}. Generating those for a field with PNGs, however, requires careful transformations of the Gaussian field. The initial conditions of \textsc{Quijote-Png}, which we use in this work for our PNG simulations, are generated using the methodology of \citet{Scoccimarro2012PNGs}. We briefly summarize this process below for the four different PNGs (see \cite{Chen2010PNGReview, Achucarro2022InflationReview} for reviews on inflation-driven PNGs) we consider in this work, as described in \citet{Coulton2022QuijotePNG}.

In general, given some Gaussian initial conditions for the gravitational potential, $\phi(\mathbf{x})$, or its Fourier equivalent $\phi(\mathbf{k})$, we can generate a field, $\Phi$, with a chosen bispectrum as
\begin{equation}\label{eqn:BispectrumGenerator}
    \Phi(\mathbf{k}) = \phi(\mathbf{k}) + \int \fNL [\delta_D] K(\mathbf{k}_1, \mathbf{k}_2)\phi(\mathbf{k}_1)\phi(\mathbf{k}_2)d^3k_1d^3k_2\,,
\end{equation}
where $[\delta_D] = \delta_D(\mathbf{k} - \mathbf{k}_1 - \mathbf{k}_2)$ is the Dirac delta function enforcing momentum conservation and $K(\mathbf{k}_1, \mathbf{k}_2)$ is a coupling kernel that contains information about the chosen bispectrum. By choosing different $K(\mathbf{k}_1, \mathbf{k}_2)$, we can generate density fields with different PNGs.

The bispectrum of the field $\Phi(\mathbf{k})$ defined in Equation \ref{eqn:BispectrumGenerator} above is
\begin{equation}
    B_\Phi = 2\fNL K(\mathbf{k}_1, \mathbf{k}_2) P_{\Phi, 1} P_{\Phi, 2} + \text{cyc.}
\end{equation}
Given a particular bispectrum template, one can find the coupling kernel, $K(\mathbf{k}_1, \mathbf{k}_2)$, that transforms the Gaussian field so as to imprint a chosen bispectrum. In this work, we focus on the same four PNG templates used in \citet{Coulton2022QuijotePNG}.

\textbf{First, the local-type} $\boldsymbol{\fNLLoc}$ is the most well-studied $\fNL$ in the context of LSS and of simulation-based studies. This is largely due to the simplicity in generating the associated initial conditions, which can be done entirely in real-space. The bispectrum of the field goes as,
\begin{equation}
    B^\Loc_{\Phi}(\kOne, \kTwo, \kThree) = 2\fNLLoc P_\Phi(\kOne)P_\Phi(\kTwo) + 2 \,\text{perms.}\,,
\end{equation}
and the field itself can be generated easily by adding the square of the Gaussian field to the linear term,
\begin{equation}\label{eqn:fnlLocalreal}
    \Phi^\Loc(\mathbf{x}) = \phi(\mathbf{x}) + \fNLLoc\big[\phi(\mathbf{x})^2 - \langle \phi(\mathbf{x})^2 \rangle\big]\,,
\end{equation}
where the ensemble average must be subtracted out to enforce that the field, $\phi(\mathbf{x})^2$, has zero mean and is thus a purely perturbative field that does not alter the mean value of $\Phi^\Loc(\mathbf{x})$. The bispectrum of this model peaks at ``squeezed'' configurations, $\kOne \ll \kTwo, \kThree$, and can be generated by the presence of a second, light scalar field during inflation, often called a ``curvaton'' \citep{Moroi2001Curvaton, Enqvist2002Curvaton, Lyth2002Curvaton, Sasaki2006Curvaton}. Such a bispectrum could also be generated during reheating --- a process during which inflatons decay into the standard model particles --- via a fluctuating, inhomogeneous decay rate \citep{Kofman2003Reheating, Dvali2004ReheatingPNG, Dvali2004Reheating2}.

\textbf{Second, the equilateral-type} $\boldsymbol{\fNLEQ}$ is a ``non-local'' PNG since it cannot be generated as local real-space transforms of the initial Gaussian field. It was derived in \citet[][see their Equation 3.1]{Senatore2010WMAP5pngs} and has a bispectrum of the form
\begin{align}
    B^\EQ_{\Phi}(\kOne, \kTwo, \kThree) = 6\fNLEQ \bigg[& -P_\Phi(\kOne)P_\Phi(\kTwo) +\textit{ 2 \text{perm.}} -2\bigg(P_\Phi(\kOne)P_\Phi(\kTwo)P_\Phi(\kThree)\bigg)^{2/3} \nonumber\\
    & +P_\Phi(\kOne)^{1/3}P_\Phi(\kTwo)^{2/3}P_\Phi(\kThree) + \textit{ 5 \text{perm.}}\bigg].
\end{align}
Such PNGs are generated from inflation models that have ``non-canonical'' kinetic terms, and their amplitude peaks in the limit $\kOne \approx \kTwo \approx \kThree$. This template approximates the bispectrum that arises from leading-derivative cubic interactions in the effective field theory (EFT) of inflation~\citep{2008EFToI}. Prototypical models with a non-canonical kinetic term and a subluminal sound speed include Dirac-Born-Infeld, or DBI, inflation \citep{2004Tong,Alishahiha2004DBI} or k-inflation \citep{Armend1999kinflation}.
The corresponding real-space expression at the field-level is
\begin{align}
    \Phi^\EQ(\mathbf{x}) = \phi + \fNLEQ\bigg[-3\phi^2 + 4 \partial^{-1}(\phi\partial\phi) + 2\nabla^{-2}(\phi \nabla^2\phi) + 2\nabla^{-2}(\partial\phi)^2\bigg]\,.
\end{align}

\textbf{Third, the CMB Orthogonal-type} $\boldsymbol{\fNLORCMB}$ is also a non-local PNG template derived in \citet[][see their Equation 3.2]{Senatore2010WMAP5pngs}, which has a shape that is approximately orthogonal to both local-type and equilateral-type PNG. Together with the equilateral type, the two shapes cover the parameter space spanned by the two leading-derivative cubic interactions in the EFT of inflation. The template takes the form 
\begin{align}
    B^\ORCMB_{\Phi}(\kOne, \kTwo, \kThree) =  6\fNLORCMB& \,\,\bigg[-3P_\Phi(\kOne)P_\Phi(\kTwo) +\textit{ 2 \text{perm.}}-8\bigg(P_\Phi(\kOne)P_\Phi(\kTwo)P_\Phi(\kThree)\bigg)^{2/3} \nonumber\\
    & +3P_\Phi(\kOne)^{1/3}P_\Phi(\kTwo)^{2/3}P_\Phi(\kThree) + \textit{ 5 \text{perm.}}\bigg]\,.
\end{align}
which leads to the real-space expression
\begin{align}
    \Phi^\ORCMB(\mathbf{x}) = \phi + \fNLORCMB\bigg[-9\phi^2 + 10 \partial^{-1}(\phi\partial\phi) + 8\nabla^{-2}(\phi \nabla^2\phi) + 8\nabla^{-2}(\partial\phi)^2\bigg]\,.
\end{align}

\textbf{Fourth and finally, the \textit{LSS} Orthogonal-type } $\boldsymbol{\fNLOR}$ is another template derived in \citet[][see their Appendix B]{Senatore2010WMAP5pngs}, which is also orthogonal to both local-type and equilateral-type PNG like $\fNLORCMB$. When considering the squeezed limit, it is a better approximation to the true bispectrum shape --- where the true shape is determined by the EFT of inflation --- when compared to the CMB orthogonal type. 
The bispectrum here is written as
\begin{align}
    B^\OR_{\Phi}(\kOne, \kTwo, \kThree) = &\,\,6\fNLOR\bigg(P_\Phi(\kOne)P_\Phi(\kTwo)P_\Phi(\kThree)\bigg)^{2/3}\bigg[\frac{p}{27}\frac{\kOne^4}{\kTwo^2\kThree^2}+ \textit{2 perms.} \nonumber\\
    &  - \frac{20p}{27}\frac{\kOne\kTwo}{\kThree^2} + \textit{2 perms.} 
    - \frac{6p}{27}\frac{\kOne^3}{\kTwo\kThree^2} + \textit{5 perms.} \nonumber\\
    & + \frac{15p}{27}\frac{\kOne^2}{\kThree^2} + \textit{5 perms.} + \bigg(1 + \frac{9p}{27}\bigg)\frac{\kThree^2}{\kOne\kTwo} + \textit{2 perms.} \nonumber\\
    & + \bigg(1 + \frac{15p}{27}\bigg)\frac{\kOne}{\kThree} + \textit{5 perms.} - \bigg(2 + \frac{60p}{27}\bigg)\bigg]\,,\\\nonumber
\end{align}
where the constant $p$ is given by
\begin{equation}
    p = \frac{27}{-21 + \frac{743}{7(20\pi^2 - 193)}}\,.
\end{equation}
The real-space expression for this template is lengthy and so instead of reproducing it here, we direct readers to Equation A11 of \citet{Coulton2022QuijotePNG}.

The ``non-local'' PNG amplitudes --- $\fNLEQ, \fNLOR, \fNLORCMB$ --- depend on the self-coupling of the inflationary perturbations. There is a natural theoretical threshold for these PNG at $\fNL\sim 1$. If $\fNL$ is much larger than unity, then the inflationary theory is favored to be strongly coupled, which in turn disfavors the simplest single-field, slow-roll model. Thus, constraining these $\fNL$ would have profound implications for understanding the physics of inflation. Given the formalism of Equation \ref{eqn:BispectrumGenerator}, one could define other templates as well, each corresponding to a different bispectrum signature that probes different interactions in the inflationary field. However, a number of models will have bispectrum shapes that overlap either/both of the local and equilateral PNG templates. The inclusion of two more templates in this work --- $\fNLOR$ and $\fNLORCMB$ --- further expands the breadth of models we can probe.

Note that all templates above are designed to induce a specific \textit{bispectrum} in the initial density field. These templates may induce additional, unintended corrections to the power spectrum, trispectrum (the Fourier space version of the 4-point correlation function), and higher-order spectra. \citet[][see their Figure 1]{Coulton2022QuijotePNG} show that the impact of $\fNL$ on the primordial power spectra is negligible --- which is by construction as the method requires corrections to the power spectrum to be subdominant \citep{Scoccimarro2012PNGs} --- while \citet[][see their Appendix A]{Jung2023fNLHMFQuijote} also show that the templates generate no unphysical trispectra. In summary, all the signals we study further below are physical and not artifacts of the PNG generation procedure.

\subsection{Simulating DES and LSST skies}\label{sec:MockMaps}

\begin{figure}
    \centering
    \includegraphics[width = 0.5\columnwidth]{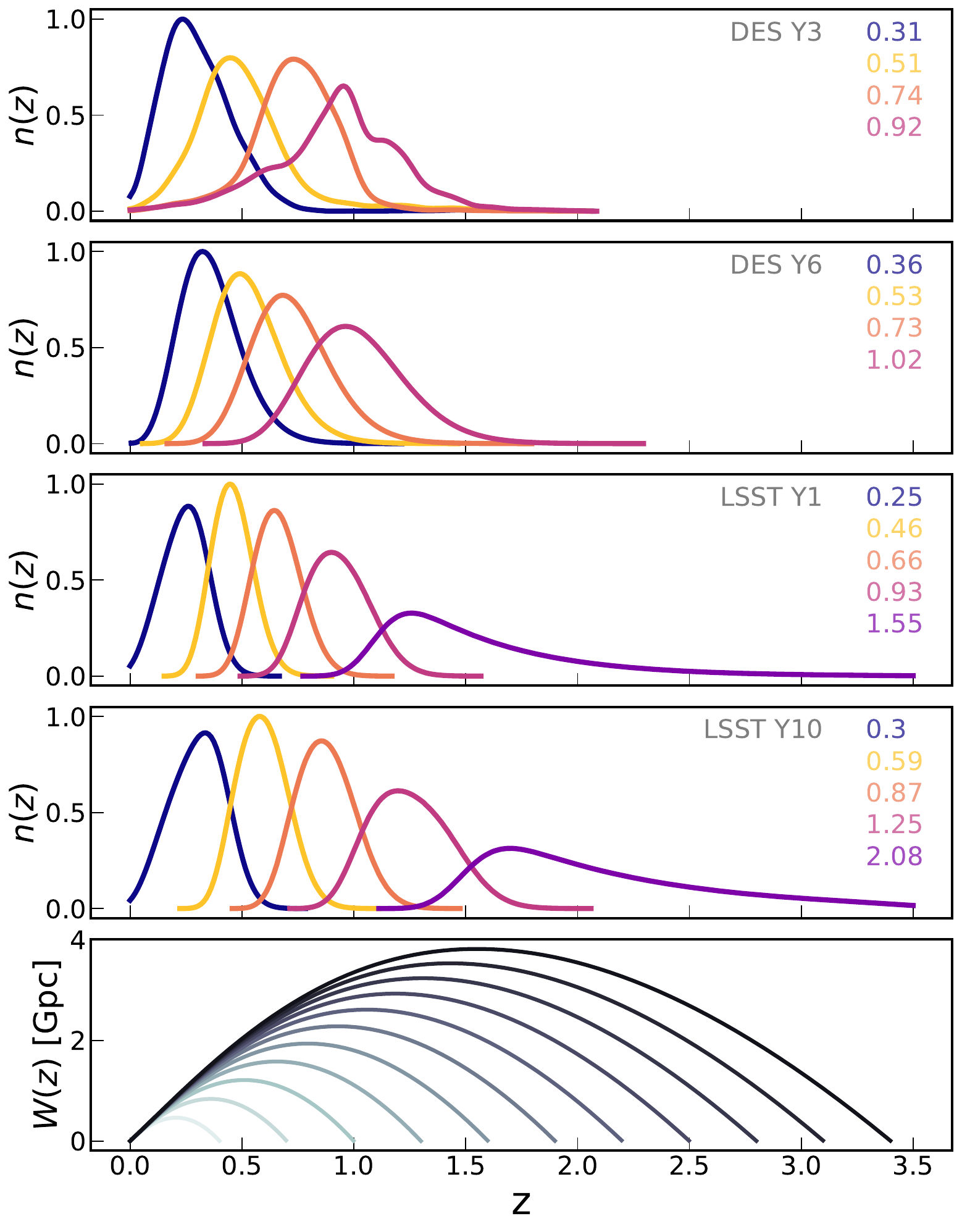}
    \caption{The redshift distribution of the signal in each tomographic bin for each survey. The DES Y3 distributions are from \citet{Myles2021PhotoZ}, and have been smoothed with a narrow Gaussian kernel for visualization purposes only. The colored numbers are the mean redshift of each bin. The bottom panel shows the quantity, $W(z_j) = \chi_j(\chi_s - \chi_j)/({a(z_j)\chi_s})$ as used in Equation \ref{eqn:convergence_definition}, with the different colors showing different $z_s$. The lines terminate at the redshift of the source plane, $z_s$. LSST has a tail to high redshifts that is likely unrealistic, but Figure \ref{fig:RedshiftDep} below shows our results are insensitive to the presence of this tail.}
    \label{fig:Nz_Wkernel}
\end{figure}

In this work, we are interested in the impact of $\fNL$ as observed by a weak lensing survey. For this, we must construct lensing maps. This can be done by using the density fields of N-body simulations to create lensing convergence fields, and then post-processing these fields to match the observed data. Various aspects of these procedures have been utilized in existing analyses/forecasts of weak lensing data \citep{Fluri2019DeepLearningKIDS,  Zurcher2021WLForecast, Fluri2022wCDMKIDS, Gatti2022MomentsDESY3, Zurcher2022WLPeaks, Gatti2023SC, Anbajagane2023CDFs}. The two surveys we focus on are: (i) the Dark Energy Survey \citep[DES,][]{DES2005}, which is an optical imaging survey of 5,000 deg$^2$ of the southern sky, and is currently the largest precision photometric dataset for cosmology. The Year 3 data products and cosmology results are available \citep{Sevilla2021Y3Gold, DES2022Y3}, while the legacy Year 6 dataset is not yet available at the time of publishing this work. (ii) the Rubin Observatory Legacy Survey of Space and Time (LSST), which is a 14,000 deg$^2$ survey that probes higher redshifts, and is the successor to current weak lensing surveys. We detail below the exact steps in our forward modeling procedure to make lensing fields corresponding to these surveys:

\textbf{Constructing lensing convergence shells.} The main simulation product used in this work is lightcone shells of the particle counts, which are a set of two-dimensional, \textsc{HEALPix} maps of projected particle counts at different redshifts. The particle count in pixel $i$ is trivially converted into the overdensity field as
\begin{equation}\label{eqn:particle_to_density}
    \delta^i = N^{i}_{\rm p}/\langle N_{\rm p} \rangle - 1,
\end{equation}
where $N^{i}_{\rm p}$ is the number of particles in pixel $i$, and the average is computed over all pixels in the shell. The density shells can then be converted into the convergence $\kappa$ using Equation \ref{eqn:convergence_definition}, after converting the integral over redshift into a discrete sum over lightcone shells.

\textbf{Source galaxy redshift distributions.} Once we have convergence shells at different redshifts, we construct the convergence within the different \textit{tomographic bins} of each survey. This binned convergence is computed as a weighted average, where the weights are the source galaxy $n(z)$ of the chosen bin and survey,
\begin{equation}
    \kappa^A(\nhat) = \sum_{j = 1}^{N_{\rm steps}} n^A(z_j)\kappa(\nhat, z_j)\Delta z,
\end{equation}
where $\kappa^A$ is the true convergence of a tomographic bin, $A$. The $n(z)$ is obtained from the following: for DES Year 3 we use the results from \citet{Myles2021PhotoZ}, for DES Year 6, LSST Year 1 and Year 10 we use the same $n(z)$ utilized in \citet[][see their Table 1]{Zhang2022CMBLSS}. The LSST modeling in that work follows the baseline analysis choices of \citet[][see their Appendix D2.1]{LSST2018SRD}. The redshift distributions for DES Y6, and LSST Y1 and Y10 are parameterized as,
\begin{equation}
    \frac{dN}{dz} \propto z^2\exp\bigg[-\bigg(\frac{z}{z_0}\bigg)^\alpha\bigg],
\end{equation}
with parameters given in Table \ref{tab:SurveySpecs}. Once the $n(z)$ of the full survey is defined, we split it into 4 (5) tomographic bins for DES Y6 (LSST Y1/Y10) of equal number density. Each bin is then convolved with a Gaussian of width given by the photometric redshift uncertainty, also quoted in Table \ref{tab:SurveySpecs}. The final $n(z)$ for each survey is shown in Figure \ref{fig:Nz_Wkernel}. The DES Y6 distribution is non-zero only between $0.2 < z < 1.3$, following \citet[][see their Table 1]{Zhang2022CMBLSS}. The LSST distributions are cut at $z < 3.5$. The fifth LSST bin (purple line) has a tail to high redshifts that is likely unrealistic. However, we will show below (in Figure \ref{fig:RedshiftDep}) that this bin has a negligible impact on our final constraints.

\begin{table}
    \centering
    \begin{tabular}{c|c|c|c}
        Survey & $(z_0, \alpha)$ & $n_{\rm gal}/{\rm arcmin}^2$ & $\sigma_z$ \\
        \hline
        \hline
        DES Y6 & (0.13, 0.78) & 9 & 0.1(1 + z)\\
        LSST Y1 & (0.13, 0.78) & 10 & 0.05(1 + z)\\
        DES Y10 & (0.11, 0.68) & 27 & 0.05(1 + z)\\
        \hline
    \end{tabular}
    \caption{The redshift distribution and source galaxy number density assumed for the upcoming surveys. All numbers are taken from \citet[][see their Table 1]{Zhang2022CMBLSS}.}
    \label{tab:SurveySpecs}
\end{table}

\textbf{Constructing lensing shear shells.} Weak lensing surveys do not directly observe the convergence, $\kappa$, as their main observable is the galaxy shapes that are tracers of the shear field, $\gamma$.\footnote{Formally, the galaxy ellipticities trace the reduced shear, $e = \gamma/(1 - \kappa)$. See Section \ref{sec:ModelChallenge} for more details.} The shear and convergence field can be transformed between each other using the Kaiser-Squires (KS) transform \citep{Kaiser1993KS}, implemented in harmonic space as
\begin{equation}\label{eqn:Kappa2Shear}
    \gamma^{\ell m}_E + i\gamma^{\ell m}_B = -\sqrt{\frac{(\ell + 2)(\ell - 1)}{\ell(\ell + 1)}} \bigg(\kappa^{\ell m}_E + i\kappa^{\ell m}_B \bigg),
\end{equation}
where $X_{\{E, B\}}$ are the E-mode and B-mode (or Q and U polarizations, in \textsc{HealPix} notation) of the field.

\textbf{Intrinsic alignments (IA).} Even in the absence of weak lensing, the observed galaxy shapes will have non-zero correlation due to the presence of a tidal gravitational field aligning the galaxy orientations. This effect is called intrinsic alignments \citep{Troxel2015IAReview, Lamman2023}. Under perturbation theory, the leading-order contribution of this effect is
\begin{equation}\label{eqn:IA}
    \kappa_{\rm IA}(\nhat, z) = -\frac{A_{\rm IA} \rho_{\rm crit, 0} \Omega_{\rm m}}{D_+(z, \Omega_{\rm m})}  \bigg(\frac{1  + z}{1 + 0.6}\bigg)^{\eta_{\rm IA}} \delta(\nhat, z),
\end{equation}
where $\kappa_{\rm IA}(\nhat, z)$ is the convergence signal corresponding to IA, $\delta(\nhat, z)$ is the density field, $A_{\rm IA}$ is the amplitude of the IA effect, $\rho_{\rm crit}$ is the critical density of the universe at $z = 0$, $\Omega_{\rm m}$ is the fraction of matter energy density at $z = 0$, $D_+$ is the linear growth function normalized to $D_+(z = 0) = 1$, and $\eta_{\rm IA}$ is the redshift scaling. Both $A_{\rm IA}$ and $\eta_{\rm IA}$ are free parameters, and will be referred to as IA parameters. The convergence field with IA is then simply $\kappa \rightarrow \kappa + \kappa_{\rm IA}$. This parameterization of IA is called the non-linear linear alignment (NLA) model, named so because it is the ``linear'', or first-order, IA correction but uses the ``non-linear'' density field. We take the first 100 simulations from the fiducial runs to include the effects of IA, and these simulations will be used to take derivatives of our data-vectors with respect to IA parameters. In specific, we create four different variations with $\eta_{\rm IA}^+ = -1.6$, $\eta_{\rm IA}^- = -1.8$ and $A_{\rm IA}^+ = 0.8$, $A_{\rm IA}^- = 0.6$. These are variations of $0.1$ around the fiducial values of $A_{\rm IA} = 0.7$ and $\eta_{\rm IA} = -1.7$ as chosen in \citet[][see their Table 2]{Krause2021Methods}.\footnote{Note that \citet{Krause2021Methods} use a more sophisticated IA model, called TATT (see Section \ref{sec:ModelChallenge}), which has additional terms beyond the NLA model of Equation \ref{eqn:IA}. We take the fiducial values of their $a_1$ and $\eta_1$ parameters, which correspond to $A_{\rm IA}$ and $\eta_{\rm IA}$ in Equation \ref{eqn:IA}.} Through these four runs, we compute the derivatives of the data-vector with respect to the IA parameters, and marginalize over these parameters in the analyses to follow. Other, more sophisticated parameterizations of the IA effect also exist, and we discuss our IA modeling approach in Section \ref{sec:ModelChallenge}.

\textbf{Shape noise.} Once the two shear fields are generated, we add the relevant shape noise in real space. In most cases, the forward-modelled field includes Gaussian shape noise with a standard deviation given as
\begin{equation}\label{eqn:shapenoise}
    \sigma_\gamma = \frac{\sigma_e}{\sqrt{n_{\rm gal}A_{\rm pix}}},
\end{equation}
where $n_{\rm gal}$ is the source galaxy number density, and $A_{\rm pix}$ is the pixel area for a given map resolution. All maps in this work use $\texttt{NSIDE}=1024$, corresponding to a pixel resolution of $3.2 \arcmin$. The per-galaxy shape noise is taken to be $\sigma_e = 0.26$. This technique is utilized for the DES Y6 and LSST Y1/Y10 fields. For DES Y3 we use a different technique given the weak lensing catalogs are already available. In this case, we use the public galaxy shape catalog\footnote{\url{https://des.ncsa.illinois.edu/releases/y3a2/Y3key-catalogs}} and rotate all galaxy ellipticities to remove any spatial correlation in the shapes. The rotated shape catalog is used to make a shear field, where each pixel value is the weighted average of the galaxy ellipticities in that pixel. This is the same technique used by other works to estimate the DES Y3 shape noise field \citep{Gatti2022MomentsDESY3, Gatti2023SC, Anbajagane2023CDFs}.

\textbf{Survey Mask \& Mass map construction.} The noisy shear field --- which is the sum of the true shear fields and the shape noise fields --- is then masked according to the survey footprint. For DES Y3 and Y6 we use the provided survey mask in the Y3 data release \citep{Sevilla2021Y3Gold}. For LSST Y1/Y10, we divide the sky into three equal-area cutouts of roughly 14,000 deg$^2$ each, which is the expected area coverage for Y10 and $\approx 15\%$ larger than the expected area for Y1. The noisy shear maps are converted to convergence using Equation \ref{eqn:Kappa2Shear}. We only use the resulting E-mode field, $\kappa_E$, for our analyses. This follows the same procedures used in \citet{Chang2018MassMap, Niall2021MassMap}. The B-mode field is non-zero in this case as the presence of a mask transfers some fraction of power (and thus cosmological information) from E-modes to B-modes. The loss of Fisher information due to this leakage can be tested by including the B-mode maps in the analysis. This inclusion will double the data-vector size --- assuming we extend the data-vector only with replications of all measurements on the B-mode field and do not also consider cross-correlations between the E-mode and B-mode fields --- which will lead to poorer numerical convergence of the final constraints. Thus, we do not test the loss in Fisher information due to this leakage.

\textbf{To summarize,} lensing convergence maps are constructed from the raw particle number count maps. The $n(z)$ distributions for a given survey are used to obtain the convergence map in a given tomographic bin. IA effects are added if desired. This convergence map is converted to shear maps, the relevant shape noise is added, the relevant survey mask is applied, and then the noisy shear maps are converted back to a noisy convergence map. The set of procedures listed above is the standard approach for forward modeling the lensing field \citep[\eg][]{Zurcher2021WLForecast, Zurcher2022WLPeaks, Gatti2022MomentsDESY3, Anbajagane2023CDFs}. Thus, our final convergence maps will be an accurate representation of the survey data.

Some known effects have been left out of our forward modeling procedure above: mean redshift uncertainties \citep[\eg][]{Myles2021PhotoZ}, multiplicative bias \citep[\eg][]{MacCrann2022ImsimsY3}, clustering of source galaxies \citep[\eg][]{Krause2021Methods, Gatti2020Moments, Gatti2023SC}, and reduced shear \citep[\eg][]{Krause2010ReducedShear, Gatti2020Moments}, to name a few. This choice has been made for simplicity, and because these factors are not expected to change the Fisher information constraints by a notable amount; either because the effect does not include a nuisance parameter to marginalize over (\eg reduced shear) or is an effect with accurate enough calibration such that the nuisance parameter marginalization has a negligible effect on the final constraints (\eg mean redshift, multiplicative bias). Nevertheless, we discuss the effects of these components in more detail in Section \ref{sec:ModelChallenge}.

\section{Higher-order statistics}\label{sec:HOS}

As mentioned prior, if a cosmological field can be defined by the covariance between different pixels, then the field's only degree of freedom is its power spectrum or 2-point correlation function. Thus the latter are the ``optimal'', i.e. lowest noise, estimators to capture this information and maximize constraints on any physics that imprints onto this field. In this work, the PNG signal of interest is inherently non-Gaussian. Therefore, our chosen summary statistic must extend beyond the 2-point function to capture the relevant information. This information is often denoted ``non-Gaussian'' or ``higher-order'' information.

In this section, we describe the different summary statistics employed in this work to capture the non-Gaussian component of the field. These include (i) the moments, which have been utilized for cosmological constraints in DES Y3 \citep{Gatti2020Moments, Gatti2023SC}, (ii) the CDFs, which have been tested for DES Y3 data \citep{Anbajagane2023CDFs}, and (iii) the three-point function, which contains the shape/configuration information whereas the former two only contain the angle-averaged component (see Section \ref{sec:Sara3-pointIntro} for more details). There exist many more choices for a summary statistic sensitive to non-Gaussian information, such as the wavelet phase harmonics, scattering transforms, mass-aperture moments, integrated 3-point functions, skew-spectrum, and more  \citep{Gruen2018DensitySplitY1,Friedrich2018DensitySplit, Gatti2020Moments, Allys2020WPHandLSS, Boyle2021MatterPDF, Cheng2021WeakLensingST, Halder2021Integrated3ptShear, Secco2022MassAp, Munshi2021, Boyle2023CGFlensing}. We choose the moments as they have been applied to data to extract cosmology, and choose the CDFs (which have been tested on data) and 3-point function as they are theoretically motivated extensions to the moments approach.

\subsection{Moments}\label{sec:MomentsIntro}

The moments of the field provide an efficient way to summarize the information from different orders. They are computed as 
\begin{equation}\label{eqn:Moments}
    \langle \kappa^{(1)}\kappa^{(2)}\ldots \kappa^{(N)}\rangle(\theta) = \frac{1}{N_{\rm pix} - 1}\sum_{i=1}^{N_{\rm pix}}\kappa^{(1)}_i\kappa^{(2)}_i\ldots \kappa^{(N)}_i\,,
\end{equation}
where $\kappa^{(j)}$ is the convergence field of the $j^{\rm th}$ tomographic bin, $N_{\rm pix}$ is the number of pixels in the survey footprint. In all cases,  $\langle\kappa^{(j)}\rangle = 0$ is enforced and the scale dependence on $\theta$ is obtained by making measurements after smoothing the fields with a harmonic-space tophat filter,
\begin{equation}\label{eqn:tophat}
    W_\ell(\theta) = 2\frac{j_1(\ell \theta)}{\ell\theta}\,,
\end{equation}
where $\theta$ is the tophat radius in radians. We use ten bins of $\theta$, between $3.2\arcmin < \theta < 200\arcmin$, which follows the analysis choices of \citet{Gatti2022MomentsDESY3}. All fields are smoothed by the same $\theta$. Varying this choice so each of the N convergence fields has a different smoothing scale can probe additional information in the field. We prioritize using the same choices as \citet{Gatti2022MomentsDESY3}, who used a single $\theta$. The quantity $\kappa_i$ in Equation \ref{eqn:Moments} is the same as the $\kappa(\nhat)$ defined in equation \ref{eqn:convergence_definition}, but with the continuous direction $\nhat$ now replaced by a discrete one defined by pixels in the \textsc{Healpix} map.

The Nth moment is sensitive to information from the N-point correlation function; the 2nd moment depends on the 2-point function, while the third moment depends on the 3-point function. In specific, the moments depend on the \textit{volume-integrated} N-point functions of the same order. This means that any dependence of the correlation on the specific shape of the N-point function is integrated over when measuring the moments. In this work, we use the 2nd, 3rd, 4th, and 5th moments. \citet[][see their Figure 6]{Anbajagane2023CDFs} measure this set of moments on DES Y3 data and find the 5th moment is consistent with no cosmological signal. However, we include it in this work as the improved precision of an LSST Y10-like dataset could enable the extraction of cosmological signals at this order. Including the sixth moment will double the length of the existing data-vector in return for marginal improvements in the Fisher information, and so we do not consider it.

The second and third moments of the field have been validated extensively to determine their robustness as a summary statistic \citep{Gatti2020Moments} and this has led to their use in DES Y3 to constrain cosmology parameters \citep{Gatti2022MomentsDESY3}. This is in contrast to the other two statistics we consider here, which are not yet validated at the rigor required for extracting lensing-based constraints.

\subsection{Cumulative Distribution Function (CDF)}\label{sec:CDFsIntro}

The CDFs are a statistic closely connected to the moments as they are both different representations of the same distribution of lensing convergence, $P(\kappa)$. The CDF measurements are defined as,
\begin{align}\label{eqn:CDFMeasurement}
    \CDF(\theta, \nu) & = P(\kappa^{(1)} > \nu, \kappa^{(2)} > \nu, \ldots \kappa^{(N)} > \nu) \nonumber\\
    & = \frac{1}{N_{\rm pix}}\sum_{i=1}^{N_{\rm pix}} \Theta(\kappa^{(1)}_i - \nu)\Theta(\kappa^{(2)}_i - \nu) \ldots \Theta(\kappa^{(N)}_i - \nu)\,,
\end{align}
where $\Theta(\kappa - \nu)$ is the Heaviside step function, which takes values of 1 if $\kappa - \nu > 0$ and 0 otherwise, and $\langle\kappa^{(j)}\rangle = 0$ is enforced similarly to the moments measurement. The quantity $\CDF(\theta, \nu)$ captures what fraction of the field, smoothed on a scale $\theta$, is above a threshold $\nu$. The $\nu$ have the same units as $\kappa$ and are dimensionless quantities. The choice of smoothing scales is the same as that of the moments, $3.2\arcmin < \theta < 200\arcmin$, and for each scale, we use 7 thresholds $\nu \in \{-20 , -6, -2,  0  ,  20,  6,  20\} \times 10^{-3}$. \citet{Anbajagane2023CDFs} determined these thresholds through an approximate optimization procedure for DES Y3 data. We employ the same for simplicity and do not explore survey-specific thresholds.  While we refer to these measurements as the CDFs, they are formally the CDF ``complement'' as the traditional CDFs are defined as the fraction of a distribution \textit{below} a certain threshold, and not above the threshold as we do here. Once again the scale dependence of the measurement enters through smoothing the field, $\kappa^{(j)}$, with a tophat filter; the same filter as defined in equation \ref{eqn:tophat}. We will consider measurements of up to the 3-field CDFs, so $N \in \{1, 2, 3\}$.

As mentioned prior, the CDFs are intimately connected to the moments. In particular, for a given threshold $\nu$, the measurement in Equation \ref{eqn:CDFMeasurement} is sensitive to moments of all orders. In the limit where the CDFs are computed with arbitrarily high number of thresholds, and the moments are computed to arbirarily high orders, the two constraints will match.\footnote{This is not formally true for every field as certain distributions --- with the log-normal being the most well-known --- cannot be represented using a finite combination of their moments. In the practical limit with noise, however, this matching is possible for every field.} \citet{Banerjee2023TracerFieldkNN} formally derive that the CDFs contain volume-integrals of all N-point correlation functions. As a consequence, the CDFs do not contain any shape/configuration information, as is the case with the moments. Naively, this would suggest the moments and CDFs cannot distinguish contributions from the different $\fNL$ types. However, we will show in Section \ref{sec:IsolatingInfo} that each $\fNL$ has a different scale- and redshift-dependence that enables distinguishing between them even without the shape/configuration information. 

\citet{Anbajagane2023CDFs} computed the impact of various lensing-based systematics on the CDF data-vector for DES Y3 data quality, and identified the relevant/negligible systematic effects. However, a theoretical model of the CDFs and an end-to-end validation of a CDF inference pipeline are required before this statistic can be used on data.

\subsection{3-point correlation function}\label{sec:Sara3-pointIntro}

Both the moments and the CDFs probe information to higher orders but only do so for the angle-averaged correlations; they contain volume integrals of the N-point functions rather than the functions themselves. For example, if the field contains a 3-point function that is non-zero only when the three points are equidistant (i.e. they form an equilateral triangle) then the moments and CDFs measurements detailed above would be unable to distinguish this from a field where a different type of triangle is the sole contributor. As an alternative, we consider the full 3-point function which includes all of this shape/configuration dependence. Inflationary signatures are often constrained using the galaxy bispectrum \citep{Cabass2022SingleFieldBOSS, Damico2022BossPNG, Philcox2022BossPNG} which contains all of this shape information and is particularly useful in distinguishing between the different types of $\fNL$. Since lensing convergence is a projected integral of the density field, the shape information will be less useful in distinguishing these types but is still expected to provide additional information beyond the volume-integrated statistics considered above.

Estimating the full 3-point function has a naive computational complexity of $\mathcal{O}(N^3)$, where $N$ is the number of points being correlated. In our work, these points correspond to convergence map pixels. Tree-based calculations, such as \textsc{TreeCorr} \citep{Jarvis2004TreeCorr}, achieve a computational complexity of $\mathcal{O}(N^2 \log_2N)$. While this is a significant speedup, it is not adequate for our requirements as we compute this statistic on $\mathcal{O}(10^4)$ survey realizations and for 35 (20) triplets between the tomographic bins of LSST (DES) per realization. Thus, computational speed is a necessity in allowing a statistic to be used in the simulation-based model.

An alternative approach explored by \citet{Philcox2022ENCORE, Sunseri2023Sara} is a Fast Fourier Transform (FFT)-based calculation of the 3-point function with a complexity of $\mathcal{O}(N \log_2N)$, which is the same as that of tree-based 2-point function estimators. We reproduce below the main aspects of the computational procedure for completeness and direct readers to \citet[][see their Section 2.2]{Sunseri2023Sara} for a detailed derivation.

A 3-point correlation function of a scalar field, henceforth denoted $\zeta$, is computed by counting triangles formed by different triplets of points. Thus, the function can be parametrized by the length of two sides, $r_1$ and $r_2$, and the angle between them $\phi$. A key choice in \citet{Sunseri2023Sara} is splitting the radial and angular dependence of $\zeta$ as 
\begin{equation}\label{eqn:zeta}
    \zeta(r_1, r_2, \hat{r}_1 \cdot\hat r_2) = \sum_{m=0}^{\infty} \zeta_m(r_1, r_2) e^{im \phi}.
\end{equation}
Note that this expansion pertains to any projected, 2D scalar fields, which in our case is the convergence field, $\kappa(\Vec{x})$. Thus, $\zeta$ is a \textit{projected} 3-point function, and $r_i$ are projected separations. The radial coefficients, $\zeta_m(r_1, r_2)$, can be computed as an area average over the convergence field $\kappa(\vec{x})$,
\begin{equation}\label{eqn:zeta_m}
    \zeta_m(r_1, r_2) = \frac{1}{2\pi^2}\int \frac{d^2 \vec{x}}{A}\, \kappa(\vec{x}) \,c_m(r_1, \vec{x})\, c^*_m(r_2, \vec{x}) \,,
\end{equation}
where the Fourier coefficients $c_m$ are
\begin{align}\label{eq:c_m}
    c_m(r_i; \vec{x}) \equiv \int d\phi\; e^{-i m \phi} \kappa(\vec{x} + \vec{r}_i).
\end{align}
Thus far, Equations \ref{eqn:zeta}, \ref{eqn:zeta_m}, and \ref{eq:c_m} are written as a function of distances $r_1$ and $r_2$. However, our final calculation will involve $\zeta$ computed in different radial and angular \textit{bins}. We can thereby bin the Fourier coefficients in circular annuli as,
\begin{equation}\label{eq:c_m_bins}
     c_{m}(r_i,\vec{x}) \to c_{m}^{\rm b}(\vec{x}) = \int_{r_{\rm min}^b}^{r_{\rm max}^b} \frac{dr_i}{A^b}r_i \int d\phi\,\, \kappa(\vec{x} + \vec{r}_i) e^{-im \phi},
\end{equation}
where $A^b = \pi[(r_{\rm max}^b)^2 - (r_{\rm min}^b)^2]$ is the area corresponding to the annular bin, and $r_{\rm min}^b$ and $r_{\rm max}^b$ are the minimum and maximum radius of bin $b$. With this binning, we write the $\zeta_m$ coefficients as 
\begin{align}\label{eqn:zeta_binned}
    \zeta_m(r_1, r_2) \rightarrow \zeta_m^{\mathrm{b_1} \mathrm{b_2}} = \frac{1}{2\pi^2}\int \frac{d^2 \vec{x}}{A}\, \kappa(\vec{x}) \,c^{\rm b_1}_m(\vec{x})\, c^{\rm b_2 \, *}_m(\vec{x}),
\end{align}
where $\rm b_1$ and $\rm b_2$ are bin indices. Equation \ref{eqn:zeta} shows that the 3-point function $\zeta$ can be constructed by summing over the coefficients $\zeta_m$ with some angular phase,
\begin{align}\label{eqn:zeta_final}
    \zeta^{\rm b_1, b_2}(\phi) = \sum_{m=0}^{\infty} \zeta_m^{\mathrm{b_1} \mathrm{b_2}} e^{im \phi}.
\end{align}
Equation \ref{eqn:zeta_final} is the final 3-point correlation function used in this work. The product within the sum is still a complex number, but we only keep the real component given the 3-point function, $\zeta$, describes correlations in real-space and has no imaginary component. When building a three-point cross-correlation between tomographic bins, we still compute Equation \ref{eqn:zeta_binned} and choose which triplet of bins provides the fields $\kappa(\vec{x})$, $\,c^{\rm b_1}_m(\vec{x})$ and $c^{\rm b_2 \, *}_m(\vec{x})$. If all fields come from the same bin then $\zeta$ is the auto-correlation, and if at least one field comes from a different bin then it is a cross-correlation.

The formalism above utilizes FFTs to compute the coefficients in Equation \ref{eq:c_m_bins}. This inherently assumes the field can be approximated as a flat field. The wide-field surveys we consider in this work, however, cannot be treated in such a manner and require spherical harmonics, which account for the curvature of the maps across the sky. To resolve this difference, we split the survey footprint into a set of smaller patches --- for which the flat-field approximation is adequate --- and compute $c^{\rm b}_m(\vec{x})$ separately for each patch. For each \textsc{HEALPix} pixel, we compute the quantity $\kappa(\vec{x}) \,c^{\rm b_1}_m(\vec{x})\, c^{\rm b_2 \; *}_m(\vec{x})$, as defined within the integral of Equation \ref{eqn:zeta_binned}, and average it across the survey footprint to obtain $\zeta_m^{\rm b_1, b_2}$. 

In our implementation, the patches follow a resolution of $\texttt{NSIDE} = 4$ which corresponds to a size $15 \times 15 \deg^2$. In practice, each flat field also includes an additional buffer region (of $8 \deg$) around the four sides, which alleviates edge effects during FFT calculations. We note, however, that any artifacts induced by the specific choices above (\eg patch size, buffer width) do not induce biases in the final Fisher information since these choices are applied consistently across all measurements in the analysis. This is also true of future applications to data: as long as the exact same computational procedure is performed on data --- and the simulations processed to look like data (where the latter is done to build a simulation-based model) --- the inference of any parameters will be unbiased. 

The other required choices in measuring this 3-point function are the tomographic bin combinations for which we compute $\zeta$, as well as the maximum $m$ mode used in computing Equation \ref{eqn:zeta_final}. For the former, we choose all triplet combinations of the tomographic bins, and this choice generates 20 (35) different combinations for DES (LSST). The maximum $m$ mode is $m_{\rm max} = 5$, which is the same choice made in \citet{Sunseri2023Sara}. We do not explore higher $m_{\rm max}$ given computing limitations. We compute $r_i$ in 6 logarithmic bins between $3.2\arcmin < r_i < 200 \arcmin$, and $\phi$ in 6 linear bins between $0 < \phi < \pi$. The choice of 6 bins in the former is set by computational cost. Increasing the number of bins extends the total compute time as $N_{\rm bin}^2$. The choice of 6 $\phi$ bins is because $\zeta$ is computed using 6 $m$-modes. The quantities $\zeta_m$ and $\zeta$ are related via Fourier modes, and in principle, a fine sampling in real-space is required to capture the Fourier coefficients (and vice-versa). Thus, it may be advantageous to continue with the $m$-mode coefficients without converting to $\phi$ bins. However, we perform the conversion so the final quantity is a real-space measure, the 3-point function, that is directly related to other existing measurements of the 3-point function \citep[\eg][]{Secco2022MassAp}. In addition, the conversion reduces the data-vector's memory footprint as $\zeta_m$ are complex coefficients, requiring a number for each of the real and complex parts, while $\zeta$ can only be real.

This 3-point estimator has not yet been applied to the weak lensing field. Thus an application of this statistic to data will require further validation. In principle, the $\zeta$ above is closely related to the 3-point shear correlation functions studied and tested in \citet{Secco2022MassAp} for DES Y3 data. However, in practice, the computational procedures of the two are significantly different --- for example, \citet{Secco2022MassAp} measure the \textit{shear} 3-point functions using galaxy shape catalogs, whereas we measure the convergence 3-point functions using pixelized maps --- and so an explicit validation must still be performed.

The total compute time for this 3-point estimator is roughly 10 times that of the moments (when computing 2nd to 5th moments). Given these computational demands, we limit our use of the 3-point function to one specific comparison test (see Section \ref{sec:Results} below) aimed at determining the Fisher information in the configuration/shape component of the 3-point function. We do not use this statistic in our fiducial constraints.

\section{$\fNL$ constraints from weak lensing}\label{sec:Results}

We now present the Fisher information on PNGs as probed by the weak lensing fields. In \S \ref{sec:StatisticsDep}, we determine the optimal summary statistic between the ones described in \S \ref{sec:HOS}, and in \S \ref{sec:SurveyDep}, show the results for different survey datasets (both current and upcoming). In \S \ref{sec:IsolatingInfo}, we isolate the signatures of inflation as seen in the lensing field, and identify the physical processes involved in generating such signatures. In all results below, the Fisher information is estimated as
\begin{equation}\label{eqn:Fisher}
    \textbf{F}_{ij} = \sum_{m,n}\frac{d\widetilde{X}_m}{d\theta_i}\big(\mathcal{C}^{-1}\big)_{mn}\frac{d\widetilde{X}_n}{d\theta_j},
\end{equation}
where $\frac{d\widetilde{X}_m}{d\theta_i}$ is the mean derivative of point $m$ in data-vector $X$ with respect to parameter $\theta_i$, where the mean is computed using 300 to 400 independent survey realizations depending on the survey. $\mathcal{C}^{-1}$ is the inverse of the numerically estimated covariance matrix and is computed while accounting for the Kaufman-Hartlap factor \citep{Kaufman1967, Hartlap2007},
\begin{equation}\label{eqn:invertcov}
    \mathcal{C}^{-1} \rightarrow \frac{N_{\rm sims} - N_{\rm data} - 2}{N_{\rm sims} - 1} \,\mathcal{C}^{-1}.
\end{equation}
We verify in Section \ref{appx:NumericalConvergence} that the covariance is well-converged for all summary statistics. The Kaufman-Hartlap factor is $\gtrsim 0.95$ for the 2nd and 3rd moments, the fiducial data-vector used in this work, and is $\gtrsim 0.55$ for the full 3-point function. Note that for the analysis using the 3-point function, we generate additional ``pseudo-independent'' realizations. These additional realizations are made by randomly rotating the original convergence maps, and adding a completely independent noise realization to them. For every independent realization, we make roughly two pseudo-independent ones and have a total of $16,000$ realizations. This increase in the number of realizations is required given the large length of the 3-point data-vector (when using no scale cuts). The set of $16,000$ realizations is used to compute the covariance of all data-vectors \textit{only} in the analysis comparing the configuration/shape information (Figure \ref{fig:ConfigDep}). We emphasize that all other results in this work \textit{do not} use any pseudo-independent realizations and only work with fully independent ones.

In all presentations of the Fisher information, we show both the raw estimate as well as one degraded by 40\%. The latter is used as a potentially pessimistic estimate of the Fisher information, and the specific choice of 40\% is because this degradation would correspond to a survey with half the expected survey volume (and thus, half the expected galaxy count, or $\approx40\%$ larger measurement uncertainties). As a result, all Fisher constraints below are shown as bands, with the lower (upper) limit corresponding to the raw (pessimistic) estimate.

\subsection{Dependence on summary statistic}\label{sec:StatisticsDep}

\begin{figure*}
    \centering
    \includegraphics[width = \columnwidth]{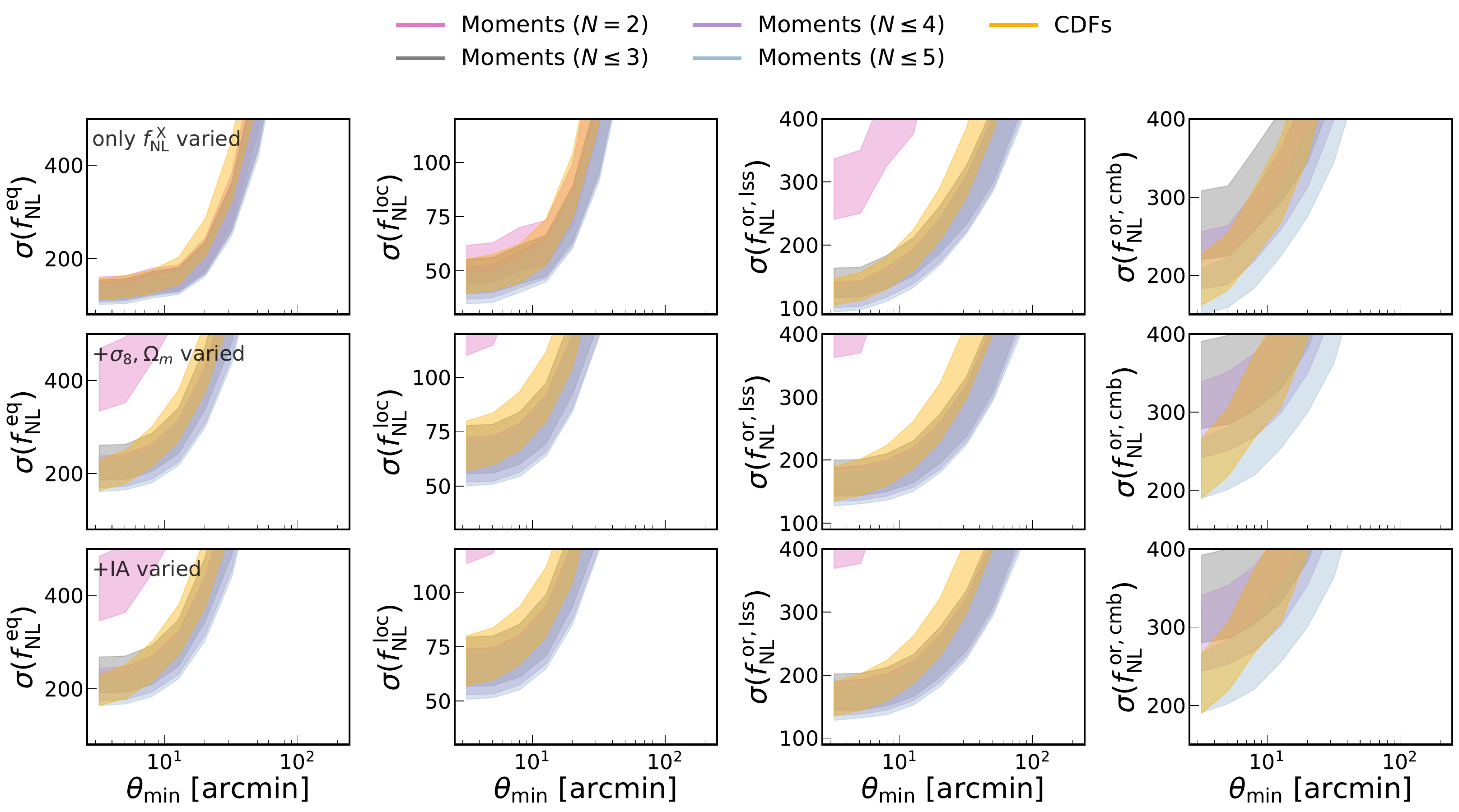}
    \caption{The Fisher information, as a function of the minimum angular scale of the data-vector, for different statistics measured on an LSST Y10-like survey. The lower (upper) bound is the Fisher (pessimistic Fisher) information, where the degradation factor of 40\% for the pessimistic case approximates constraints from half the expected survey volume. The columns correspond to different $\fNL$, and the rows progressively step from constraints varying just $\fNL$, to marginalizing over $\sigma_8$ and $\Omega_{\rm m}$, and finally to also marginalizing over the intrinsic alignment parameters, $\eta_{\rm IA}$ and $A_{\rm IA}$. The constraints for $\fNLLoc$ and $\fNLEQ$, when varying just $\fNL$, are consistent across all statistics. However, once other parameters are included in the analysis, the statistics sensitive to non-Gaussianities are clearly more favorable. The combination of the 2nd and 3rd moments does fairly similarly to all other non-Gaussian statistics considered, with the exception of analyses of $\fNLORCMB$. The 2nd moment constraints are above the range of many panels.}
    \label{fig:StatDep}
\end{figure*}

\begin{figure*}
    \centering
    \includegraphics[width = \columnwidth]{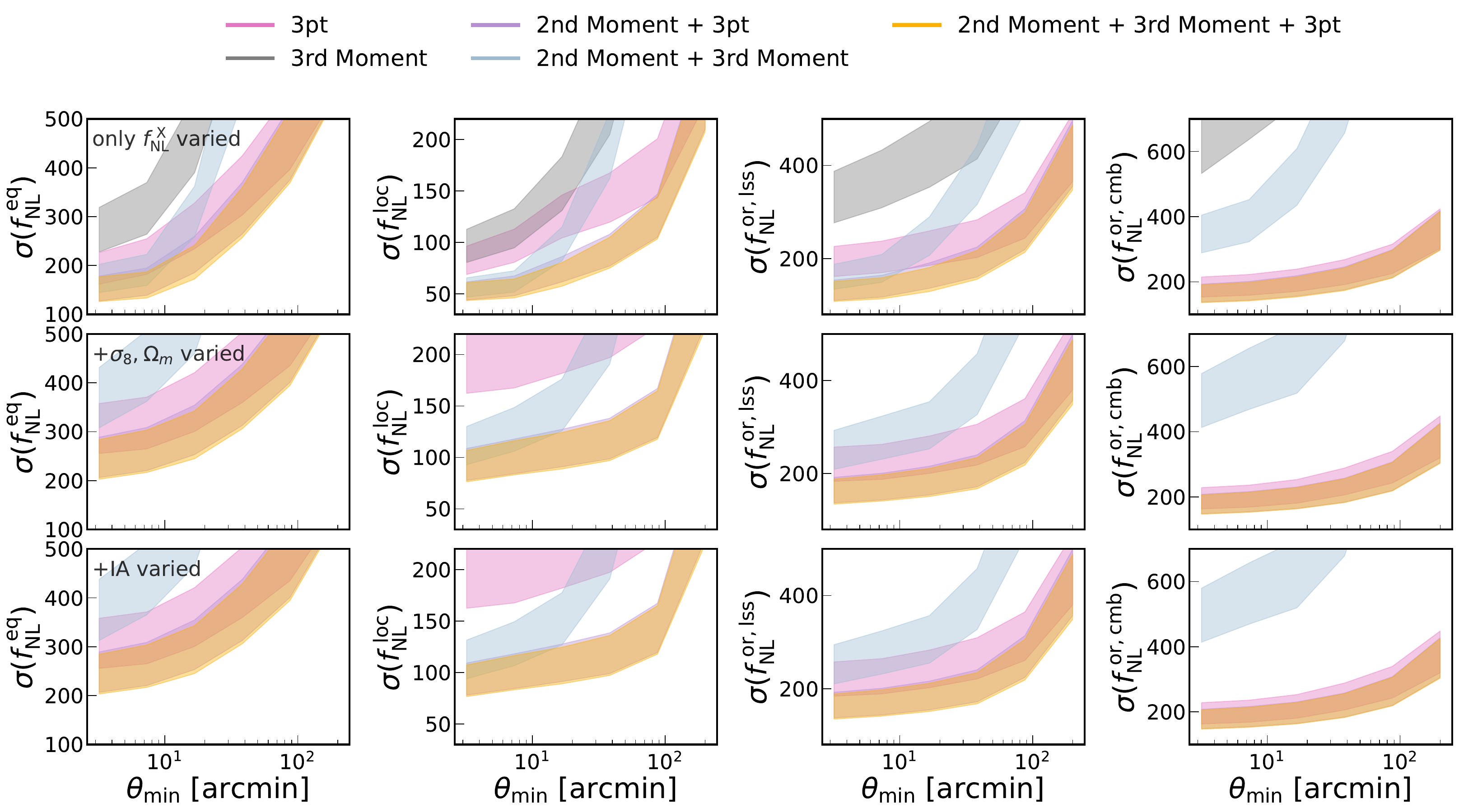}
    \caption{Similar to Figure \ref{fig:StatDep} but for the moments and the full 3-point correlation function. The purple bands in most panels overlap completely with the yellow bands and are not seen. The 3-point function constraints are considerably better than the 3rd moment constraints, indicating the configuration information found in the former (and missing in the latter) is significant. Note that the constraints for the moments are weaker here than those in Figure \ref{fig:StatDep} as all summary statistics are measured within 6 radial bins rather than the fiducial choice of 10. This is motivated by computing limitations for the 3-point function, and is discussed further in the text. The 3rd moment constraints are above the range of many panels.}
    \label{fig:ConfigDep}
\end{figure*}

Figure \ref{fig:StatDep} shows the Fisher information in different summary statistics for an LSST Y10 survey. The first row shows constraints when varying just $\fNL$ (with y-labels denoting the specific type being varied). In the second row we marginalize over $\sigma_8$ and $\Omega_{\rm m}$, and in the third row we also marginalize over the IA parameters. All results are shown as a function of minimum angular scale of the data-vector, and all discussions on scale-dependence are in Section \ref{sec:IsolatingInfo}. For $\fNLEQ$ and $\fNLLoc$, in the unmarginalized case, all statistics considered here are roughly equivalent. In particular, the 2nd moments alone are an adequate statistic, and including the higher order moments adds only marginally to the Fisher information. On non-linear scales, the late-time matter power spectra contain strong signatures from these two types of $\fNL$ \citep[][see Figure 6]{Coulton2022QuijotePNG} and can thus constrain these parameters on their own. Higher-order spectra, such as the matter bispectrum and trispectrum, will contain additional information but are also more noise-dominated than the power spectra. Hence, their contribution to the total Fisher information can be minuscule compared to the contribution from the power spectra. This behavior is seen in Figure \ref{fig:StatDep}, where the 2nd moments dominate the constraining power over the 3rd, 4th and 5th moments. For both orthogonal-type $\fNL$, the constraints from the 2nd moments alone are significantly weaker than combinations with higher-order moments. This is expected as \citet[][see Figure 1]{Coulton2022QuijotePNG} shows the non-linear power spectrum has only minimal (negligible) changes when varying $\fNLOR$ ($\fNLORCMB$).

The relevance of the higher-order information improves significantly when extending the parameter space to perform a \textit{marginalized} analysis of $\fNL$. The changes in the power spectrum due to PNGs have some overlap with changes due to gravitational evolution. Thus, marginalizing over cosmology (which is marginalizing over gravitational evolution) results in a strong degradation in the Fisher information of PNGs as probed by the 2nd moments. Including the 3rd moment significantly improves the constraints. The degeneracy-breaking from combining 2nd-order and 3rd-order information has been explored extensively in the literature \citep[\eg][]{Gatti2020Moments, Zurcher2021WLForecast, Gatti2022MomentsDESY3, Zurcher2022WLPeaks, Anbajagane2023CDFs}. They have also been explicitly shown for the PNGs we study here \citep{Coulton2022QuijotePNG}. Including the 4th and 5th moments improves the constraints slightly. The CDFs are generally similar to the combinations of 2nd and 3rd moments, and are better/worse depending on the exact analysis being performed: they are better in constraining $\fNLEQ$, worse for $\fNLLoc$, and comparable for $\fNLOR$ and $\fNLEQ$. This is generally consistent with the behaviors of the moments and CDFs found in the $w{\rm CDM}$ analysis of \citet{Anbajagane2023CDFs}. Further marginalizing over IA parameters results in minimal degradation of the $\fNL$ constraints.

\textbf{Importance of configuration/shape information.} All the statistics above have utilized the angle-averaged information in the field. These statistics involve volume integrals over the correlation functions of the field and have no sensitivity to the shape of the correlations. In Section \ref{sec:Sara3-pointIntro}, we discussed the potential information in the configuration/shape information of the field's spatial correlations. Here, we explicitly check this by computing the full 3-point correlation function and comparing its constraints to those from the 3rd moments, which have no shape information. Due to computing limitations, we only measure the 3-point functions in 6 radial bins between $3.2\arcmin < \theta < 200 \arcmin$, as opposed to the 10 bins used in the main analysis.\footnote{The calculation complexity goes as $N_{\rm r,bin}^2$ so using 10 radial bins (instead of 6) extends the compute time by a factor of 3.} To perform a fair comparison between statistics, we also remeasure the moments in 6 radial bins as well. However, we will show that this reduction in bins reduces the Fisher information on PNGs as probed by the moments. Therefore, we use the measurements of moments with 6 radial bins only for the comparison in this section and use the fiducial measurements with 10 bins for all other analyses. The moments here continue to be measured directly on the spherical sky, and do not use the patch-by-patch flat-sky approach used for the 3-point function estimator.

Figure \ref{fig:ConfigDep} shows the constraints from the 3rd moments and the 3-point function, as well as from combinations with the 2nd moments and from the combination of all three. The Fisher information for all $\fNL$ types is higher in the full 3-point function in comparison to the 3rd moment. This increase in information is still notable when varying only $\fNL$, and increases to $50\%$ improvements when marginalizing over cosmology and/or IA. The exception is $\fNLORCMB$ where the constraints are improved by a factor of 2 or more in all settings. The combination of 2nd moment and 3-point function does better than the latter alone. Adding the 3rd moment to this combination leads to no improvement; the purple and yellow bands are atop each other for most panels. This emphasizes that the 3-point function contains all the information in the 3rd moments, as expected. This behavior is consistent regardless of the set of parameters being varied in the analysis.

We have verified, using the techniques detailed in Appendix \ref{appx:NumericalConvergence}, that the marginalized $\fNL$ constraints from using the 3-point function (either on its own or combined with other statistics) are numerically converged to within 10-20\%, indicating that the constraints in Figure \ref{fig:ConfigDep} are potentially overestimated by 10-20\%. This non-convergence arises due to noise in the numerically estimated derivative. It is not related to the covariance, as we have verified the constraints change by $<1\%$ when changing the number of realizations used in estimating the covariance. The numerical convergence-based overestimate of 10-20\% is lower than the 50\% improvement mentioned above. Therefore, it is still likely that the configuration/shape measurement leads to an increase in the Fisher information of PNGs. A robust estimate of this increase will require better numerical convergence in the estimates of the derivatives.

The poorer numerical convergence of constraints from the 3-point function, compared to the percent-to-sub-percent convergence of constraints from the CDFs and the moments, can be improved by employing more independent realizations to estimate the derivatives. Note that the convergence issues have occured even after we consciously reduced the size of the data-vector by limiting the radial binning to 6 bins instead of the fiducial 10 bins. Comparisons of the moments-based constraints in Figure \ref{fig:StatDep} and Figure \ref{fig:ConfigDep} also show that this change in binning leads to significantly degraded constraints. This highlights the practical challenges in robust use of the three-point function while performing simulation-based modeling. One could compress the datavector to reduce its size and thus, the number of simulations needed to estimate the covariance. Studies on data have used SVD decompositions \citep[\eg][]{Zurcher2022WLPeaks} or the MOPED compression \citep[\eg][]{Gatti2022MomentsDESY3}. Bispectrum estimates on the CMB have also used Modal estimators to compress the data vector \citep{Fergusson2010Modal}. Given our goal of performing a true comparison between the 3-point function and the 3rd moments, we do not employ any compression. Though, we note the MOPED compression, in principle, should not degrade the Fisher constraints. In our case, this is complicated by the fact that the MOPED compression depends on our simulation-based derivatives of the data-vectors, and the noise in these derivative estimates will lead to poorer compression. We do not explore this further.

Given the discussions above on the benefits and detriments in practical implementations of each statistic, we henceforth use the combination of the 2nd and 3rd moments as the fiducial statistic for this work and for all the Fisher information analyses below.

\subsection{Fisher information in DES and LSST}\label{sec:SurveyDep}

\begin{figure*}
    \centering
    \includegraphics[width = \columnwidth]{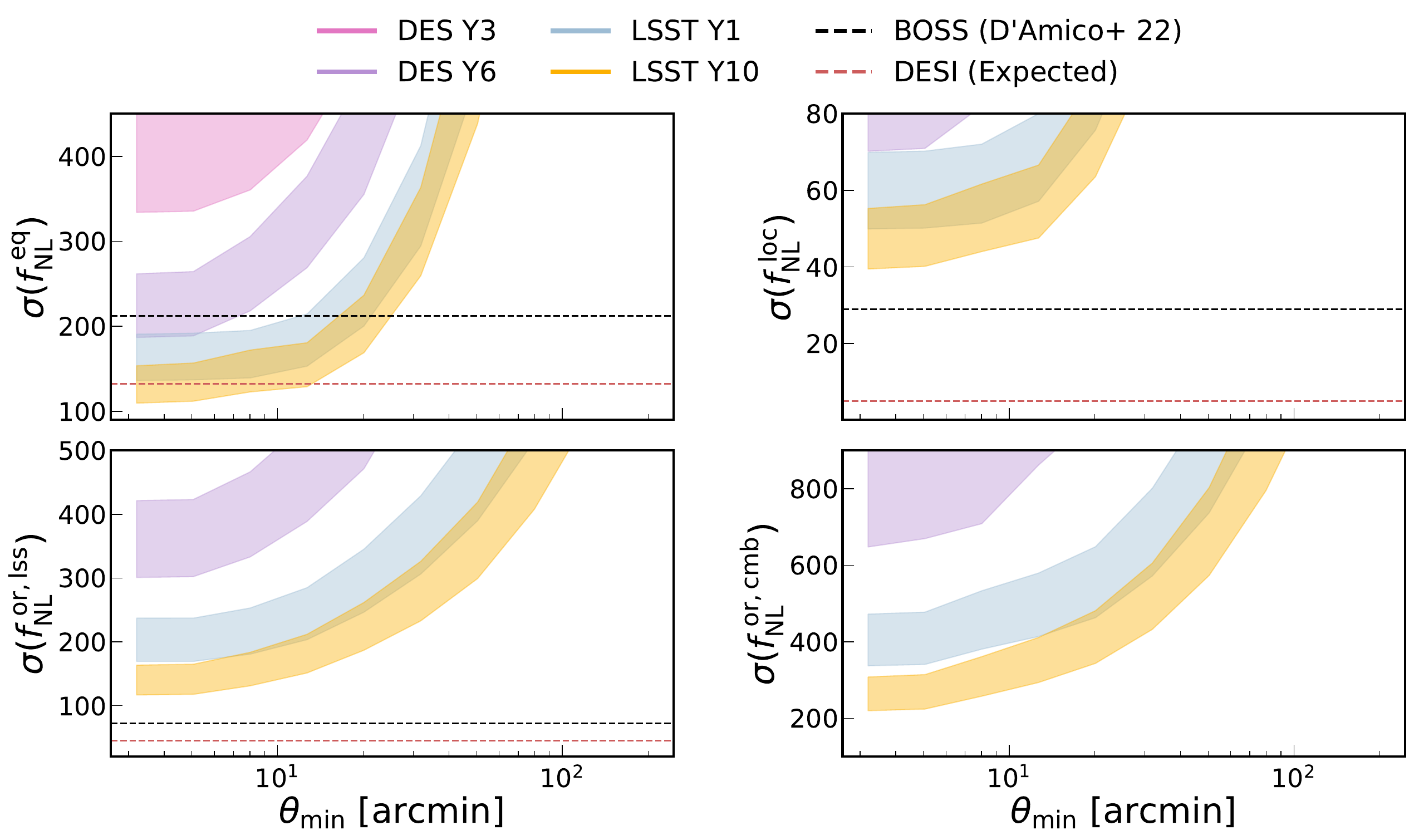}
    \caption{The Fisher information in the 2nd and 3rd moments of the lensing field, shown as a function of the minimum angular scale of data-vector, for four different survey configurations. The lower (upper) bound is the Fisher (pessimistic Fisher) information, where the degradation factor of 40\% for the pessimistic case approximates constraints from half the expected survey volume. The existing constraints from BOSS \citep{Damico2022BossPNG} are shown as the black line and the potential constraint from DESI \citep{Damico2022BossPNG, DESI2016Science} in red. The presented lensing and galaxy constraints are both consistent in fixing cosmological parameters. The DES Y6 constraints for $\fNLEQ$ are comparable to the existing BOSS constraints. The LSST Y10 constraints are comparable or potentially better than DESI for $\fNLEQ$, a factor of 2 broader for $\fNLOR$, and a factor of 8 broader for $\fNLLoc$. The DES Y3 constraints are above the range of the plots for most panels.}
    \label{fig:SurveyDep}
\end{figure*}

\begin{table}
    \centering
    \begin{tabular}{c|c|c|c|c}
        Survey & $\sigma(\fNLEQ)$ & $\sigma(\fNLLoc)$ & $\sigma(\fNLOR)$ & $\sigma(\fNLORCMB)$ \\
        \hline
        \hline
        \multicolumn{5}{c}{\textit{Fiducial Fisher information, $\theta_{\rm min} > 3.2\arcmin$}}\\
        \hline
        DES Y3 & 334 [984] & 125 [315] & 562 [850] & 1136 [1456]  \\
        DES Y6 & 187 [575] & 70 [186] & 300 [524] & 648 [905] \\
        LSST Y1 & 136 [295] & 50 [90] & 169 [234] & 337 [457]\\
        LSST Y10 & 109 [186] & 39 [55]& 116 [142] & 220 [278]\\
        \hline
        \multicolumn{5}{c}{$\theta_{\rm min} > 20\arcmin$}\\
        \hline
        DES Y3 & 530 [1279] & 203 [397] & 745[946] & 1451 [1809] \\
        DES Y6 & 355 [833] & 137 [262] & 471 [618] & 993 [1355] \\
        LSST Y1 & 200 [437] & 75 [124] & 246 [280] & 463 [555]\\
        LSST Y10 & 168 [338] & 63 [92] & 187 [195] & 343 [381]\\
        \hline
        \multicolumn{5}{c}{\textit{Galaxy correlation functions}}\\
        \hline
        BOSS (fix) & 212 & 29 & 72  & ---\\
        BOSS (vary) & 350 & --- & ---  & ---\\
        DESI (fix) & 133 & ---& 45 & ---\\
        DESI (vary) & 220 & 5 & --- & ---\\
        \hline
    \end{tabular}
    \caption{The Fisher information presented in Figure \ref{fig:SurveyDep} for the full range of scales (top) and a conservative scale cut (middle). The numbers in square brackets are constraints after marginalizing over $\Omega_{\rm m}$, $\sigma_8$, and two IA parameters. The ``BOSS (fix)'' results, analyzed at fixed cosmology, are from \citet{Damico2022BossPNG} while the ``BOSS (vary)'' results from also varying $\sigma_8$ are from \citet{Philcox2022BossPNG}. The expected constraints from DESI for $\fNLEQ$ and $\fNLOR$ are obtained by rescaling the corresponding BOSS results by 1.6 as mentioned in \citet{Damico2022BossPNG}, and for $\fNLLoc$ we take the numbers from \citet{DESI2016Science}. Lensing measurements at $\theta_{\rm min} = 3.2\arcmin$ ($\theta_{\rm min} = 20\arcmin$) correspond to power spectra at $k \in [1, 2.5]\,\, h/{\rm Mpc}$ ($k \in [0.3,0.6]\,\, h/{\rm Mpc}$), depending on the redshift bin. These are higher than the current $k_{\rm max} = 0.2 h/{\rm Mpc}$ limit of galaxy correlation analyses. The difference is more substantial if we consider the maximum scale (and not the average scale) that contributes to the lensing measurement. These scale differences are discussed further in Section \ref{sec:simlensing}.}
    \label{tab:FidConstraints}
\end{table}

The key goal of this work is extracting the Fisher information on $\fNL$ from weak lensing, and comparing it to the information in galaxy correlation functions from Baryon Oscillation Spectroscopic Survey (BOSS) or expected from DESI. Such comparisons motivate/inform the utility of weak lensing in PNG analyses. We study two versions of the DES and LSST surveys, and compare their constraints with current constraints from BOSS and expected ones from DESI. The BOSS constraints come from the analysis of \citet{Damico2022BossPNG}, while the expected DESI numbers are taken from (i) \citet{Damico2022BossPNG} for $\fNLEQ$ and $\fNLOR$, as they compute the expected constraints to be $1.6$ times higher than their BOSS constraints, and (ii) \citet{DESI2016Science} for $\fNLLoc$. Note that other works \citep{Cabass2022MultifieldBOSS, Philcox2022BossPNG} show significantly different results from \citet{Damico2022BossPNG}, primarily due to the choice of which parameters to marginalize over; the latter fixes cosmology while the former vary cosmology and other nuisance parameters. We will present results for constraint with and without such marginalization (Table \ref{tab:FidConstraints}).

Figure \ref{fig:SurveyDep} presents the Fisher information for weak lensing surveys as a function of the minimum angular scale used in the analysis. We detail our findings below for each of the four different $\fNL$ explored:

\textbf{For $\boldsymbol{\fNLLoc}$}, galaxy correlations have significantly more information than weak lensing. This is expected given the signature of $\fNLLoc$ in galaxy correlation functions is a scale-dependent galaxy bias in the 2-point function, where the bias increases towards large scales as $b \propto k^{-2}$ with $k$ being the Fourier wavenumber \citep{Dalal2008ScaleDependentBias}. Thus, this effect has a large (diverging) amplitude towards large scales and is more easily distinguished from gravitational evolution. The Fisher information in LSST (DES) is factors of 3 to 8 times lower than the information in DESI (BOSS). Thus, the benefit from including weak lensing in analyses of $\fNLLoc$ is unlikely to be from the weak lensing-only constraints and would instead be from constraints of galaxy bias parameters via the cross-correlation of lensing and galaxies. These galaxy bias parameters, numbering $\mathcal{O}(10)$ in the latest models \citep{Damico2022BossPNG, Cabass2022SingleFieldBOSS, Cabass2022MultifieldBOSS, Philcox2022BossPNG}, are required in the modeling of the correlation functions and cannot be known a priori. A more detailed discussion of this aspect can be found in Section \ref{sec:simlensing}. It is also possible that the weak lensing-only constraints enable degeneracy breaking that \textit{does} benefit the final $\fNLLoc$ constraints. The analysis in this work does not compute the Fisher information in the galaxy correlations and is unable to test this.

\begin{figure*}
    \centering
    \includegraphics[width = \columnwidth]{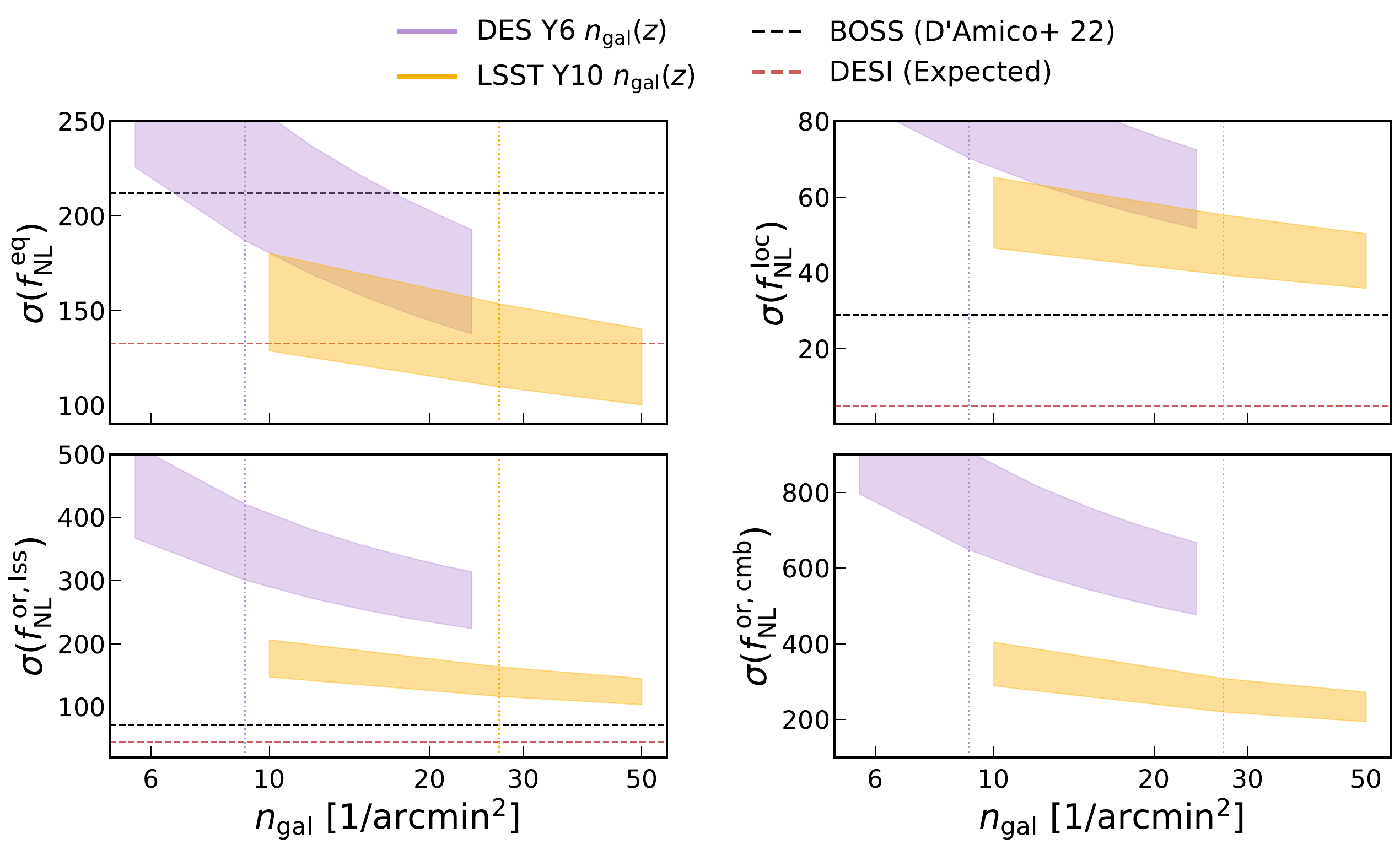}
    \caption{The Fisher information for DES Y6 and LSST Y10 measured using the 2nd and 3rd moments, as a function of source galaxy number density ($n_{\rm gal}$). The fiducial values are $n_{\rm gal} = 9 \,(27)$ for DES Y6 (LSST Y10), and are denoted by the vertical dotted line. This tests the improvement in constraints from source galaxy number density improvements alone. At fixed $n_{\rm gal}$, the difference between DES and LSST shows the difference in constraints primarily from survey area (with some impact from increasing redshift limits in LSST, see Figure \ref{fig:RedshiftDep}). For $\fNLEQ$, where weak lensing is a promising probe, the LSST survey area improves constraints by 30\% over DES Y6. The difference between constraints of LSST Y10 at $n_{\rm gal} = 10$ (close to the DES Y6 fiducial value) and LSST Y10 $n_{\rm gal} = 27$ (the LSST Y10 fiducial value) is 20-30\%.}
    \label{fig:NoiseDependence}
\end{figure*}

\textbf{For $\boldsymbol{\fNLEQ}$}, weak lensing provides constraints that are competitive with galaxy clustering. The Fisher information in LSST (DES) is similar to, and potentially better than, DESI (BOSS). Note that the difference in constraining power between DES Y3 and DES Y6 is about 60\%. The Y3 analysis uses the actual Y3 noise fields and redshift distribution/uncertainties, while the Y6 results use expected noise fields and redshift distributions. The improvement in DES Y6 over DES Y3 is estimated to be (i) a 50\% increase in galaxy sample size, which implies a 25\% increase in constraints, and (ii) an increase in the highest redshifts measured, where the amplitude of the lensing signal grows with redshift and this is expected to cause significant improvement in $\fNL$ constraints (see Figure \ref{fig:RedshiftDep}). We also find that a DES Y6 survey with the DES Y3 number density still performs better than the DES Y3 fiducial survey (see Figure \ref{fig:NoiseDependence}) which further signifies the importance of the higher maximum redshift in DES Y6 compared to DES Y3. These all signify that the 60\% improvement is reasonable.

\textbf{For $\boldsymbol{\fNLOR}$ and $\boldsymbol{\fNLORCMB}$}, the constraints from weak lensing are about 1.5 to 4 times broader than those from galaxy clustering. We only compare $\fNLOR$ as the BOSS/DESI analyses do not measure $\fNLORCMB$. The DES constraints for the former $\fNL$ type are more than a factor of 3 wider than the existing BOSS constraints, while the LSST constraints are a factor of 1.5 to 2 wider than those of DESI. Therefore, lensing can still provide valuable information in constraining these $\fNL$ amplitudes for Stage IV surveys.

\textbf{The sound speed, $\boldsymbol{c_s}$, and the EFT parameter $\boldsymbol{\widetilde{c}_3}$,} capture the two leading-derivative cubic interactions of the EFT of inflation \citep[\eg][see their Equation 1]{Cabass2022SingleFieldBOSS}. These parameters can be directly inferred from $\fNLEQ$ and $\fNLOR$ as
\begin{align}\label{eqn:EFTParams1}
    c_s & = \bigg[1 + \frac{324}{85}\bigg(1.3\fNLEQ +  8.97\fNLOR\bigg)\bigg]^{-1/2},\\[10pt]
    (c_s^{-2} - 1)\widetilde{c}_3 & = \frac{243}{10}\bigg(-0.293\fNLEQ - 7.71\fNLOR\bigg) - \frac{3}{2}(c_s^2 - 1),\label{eqn:EFTParams2}
\end{align}
where the expressions are taken from the conversions presented in Equation 5 of \citet{Cabass2022SingleFieldBOSS}. Since the constraints allow the value $c_s = 1$ (and thus $1/(c_s^2 - 1)$ is undefined) one can only constrain the combination $(c_s^{-2} - 1)\widetilde{c}_3$ rather than $\widetilde{c}_3$. Equations \ref{eqn:EFTParams1} and \ref{eqn:EFTParams2} require joint constraints on $\fNLEQ$ and $\fNLOR$, whereas we have thus far only varied one $\fNL$ at a time. Upon doing the joint analysis, we obtain the following $c_s$ constraints for DES Y6 and LSST Y10, depending on whether or not we marginalize over cosmology and IA parameters, 
\begin{align}\label{eqn:Results}
    c_s^{\rm DES \,Y6} \gtrsim 
    \begin{cases}
    0.021,& \text{unmargin.}\\
    0.011,& \text{margin.}\\
    \end{cases}\hspace{20pt}[\text{lower}\,\, 95\%]\\
    c_s^{\rm LSST \,Y10} \gtrsim 
    \begin{cases}
    0.028,& \text{unmargin.}\\
    0.016,& \text{margin.}\\
    \end{cases}\hspace{20pt}[\text{lower}\,\, 95\%]
\end{align}
where we have followed \citet{Planck2020PNGs, Cabass2022SingleFieldBOSS} in marginalizing over $\widetilde{c}_3$ and presenting constraints on just $c_s$. We stress that the results above are only rough estimates. This is because our analysis varies $\fNLEQ$ and $\fNLOR$, and not $c_s$ and $\widetilde{c}_3$ directly. Not all of the parameter space spanned by the former leads to well-defined quantities in the latter \citep{Cabass2022SingleFieldBOSS}. A more robust estimate would be obtained by running simulations that vary $c_s$ and $\widetilde{c}_3$ directly. We do not have such simulations so do not do this. If we consider the marginalized case, bounds for both DES Y6 and LSST Y10 are similar to those from the BOSS analysis of \citet{Cabass2022SingleFieldBOSS}, which finds $c_s \geq 0.013$, and the CMB analysis of \citet{Planck2020PNGs}, which finds $c_s \geq 0.021$. Both are 95\% confidence lower bounds corresponding to our marginalized case above. Given the approximate nature of our estimates above (as the simulations do not directly vary $c_s$ and $\widetilde{c}_3$) we do not interpret the comparison with more detail. Note that even though the moments integrate over shape information via the volume integral, they can still distinguish --- and thus jointly constraint --- different $\fNL$ given their different redshift and scale-dependence (see Section \ref{sec:IsolatingInfo}). \citet[][see their Appendix D]{Friedrich2020PDFBulkfNL} discuss that smoothing the density field with circular apertures is not optimal for distinguishing the different $\fNL$. Thus, varying the shape of the smoothing filter could further improve the constraints above.

\begin{table}
    \centering
    \begin{tabular}{c|c|c|c|c}
        Mom. Order & $\sigma(\fNLEQ)$ & $\sigma(\fNLLoc)$ & $\sigma(\fNLOR)$ & $\sigma(\fNLORCMB)$ \\
        \hline
        \hline
        $N = 2$ & 67 [96] & 25 [33] & 75 [95] & 164 [196] \\[2pt]
        $N = 3$ & 99 [184] & 30 [92] & 82 [178] & 128 [180] \\[2pt]
        $N = 4$ & 126 [235] & 39 [173] & 85 [171] & 121 [155] \\[2pt]
        $N = 5$ & 152 [246] & 52 [197] & 97 [165] & 121 [140] \\[2pt]
        \hline
        $N \leq 3$ & 51 [66] & 16 [17] & 39 [43] & 68 [75] \\[2pt]
        $N \leq 4$ & 40 [48] & 13 [13] & 28 [32] & 47 [55] \\[2pt]
        $N \leq 5$ & 20 [24] & 6 [7] & 13 [16] & 22 [25] \\[2pt]
        \hline
    \end{tabular}
    \caption{Constraints from different orders of the true convergence field (i.e. noiseless) for LSST Y10-like source galaxy redshift bins/distributions. We show results both for the individual moments and for combinations that successively include higher-order moments. Constraints from including up to the 5th moment improve on those of the 2nd moment alone by factors of 2 to 6. The square brackets denote constraints after marginalizing over $\Omega_{\rm m}$, $\sigma_8$, and the IA parameters. Constraints from the combination of moments are degraded by 10\% to 20\% upon marginalization, whereas those from the individual moments are impacted more.}
    \label{tab:IdealSurvey}
\end{table}

\textbf{Variation with shape noise.} We then consider variations of the DES Y6 and LSST Y10 surveys where the shape noise amplitude is artificially increased/decreased compared to the fiducial value. This tests the dependence of the constraints on the noise alone. Therefore, for this test, the source galaxy redshift distributions, the survey footprints, and all other aspects are still fixed to fiducial values for each survey. Figure \ref{fig:NoiseDependence} shows the Fisher information in each survey as a function of source galaxy number density, which is inversely related to the noise level as shown in Equation \ref{eqn:shapenoise}. Focusing on $\fNLEQ$, the constraints from a DES Y6 survey with $n_{\rm gal} = 24$ are 40\% better than those of a fiducial survey with $n_{\rm gal} = 9$. At fixed source galaxy number density, the LSST Y10 constraints are $\approx 45\%$ better than those of DES Y6. We also compute constraints for an LSST Y10-like survey but with $n_{\rm gal} = 50$, which corresponds to the number density of a potential weak lensing sample from the Roman space telescope \citep{Eifler2021Roman}. The constraints, in this case, improve only by $10\%$ for $\fNLEQ$ and $\fNLOR$. Using simple scaling arguments for the dependence of measurement noise on galaxy number density and sky area, the Fisher information is roughly proportional to $\sqrt{n_{\rm gal}}$ and $\sqrt{f_{\rm sky}}$, where $f_{\rm sky}$ is the fraction of the full sky covered by the survey. The exact scaling with $n_{\rm gal}$ can deviate from expectations depending on the scale-dependence of the signal in question.

\textbf{Interplay of different orders.} It is also useful to understand the ``true'' information at different orders of the true convergence field. Table \ref{tab:IdealSurvey} shows this, as probed by the moments. In specific, we extract the information content in the individual moments, and in their combinations. For the 4th and 5th moments, we only considered the connected piece that is obtained by subtracting out different combinations of 2nd moments or 2nd and 3rd moments, respectively. For an auto-correlation with a single field, the expression is
\begin{equation}
    \langle\kappa^4\rangle^{\rm conn} = \langle\kappa^4\rangle - 4\langle\kappa^2\rangle\langle\kappa^2\rangle,
\end{equation}
\begin{equation}
    \langle\kappa^5\rangle^{\rm conn} = \langle\kappa^5\rangle - 10\langle\kappa^2\rangle\langle\kappa^3\rangle.
\end{equation}
This modification accounts for the fact that a field with a 2nd and 3rd moment will automatically have a 4th and 5th moment. The subtraction removes such contributions and extracts the ``connected'' 4th and 5th moment. All numbers in Table \ref{tab:IdealSurvey} correspond to only connected information. The table shows that including up to the 5th moment improves the constraints from the 2nd moment-only case by factors of 2 to 6, highlighting the significant information from inflation contained in the higher-order moments. We have verified that all numbers are converged to within $1\%$. Note that Figure \ref{fig:StatDep} has already identified that in the practical limit, the 2nd and 3rd moments contain almost all of the information. Table \ref{tab:IdealSurvey} shows the impact of the higher-order information that has been lost due to the larger amplitude of noise (relative to the 2nd moment case) in these higher-order moments.

\subsection{Physical origin of $\fNL$ signatures in the weak lensing field}\label{sec:IsolatingInfo}

The discussions above have thus far established there is valuable information in the weak lensing field on PNG signatures. It is therefore imperative to identify how these signatures imprint into this field and into the data-vectors we study. Such identification will further focus our efforts on mitigating the relevant systematics and/or on selecting a data-vector that is more tuned for inflationary signatures.

\textbf{Scale dependence.} Figure \ref{fig:SurveyDep} already shows that the more non-linear scales contain the most information, and Figure \ref{fig:StatDep} and \ref{fig:ConfigDep} confirm this is true for all summary statistics we discuss here and for both the marginalized and unmarginalized constraints. Such behavior is expected as varying $\fNL$ changes the tails of the initial density distribution, which then changes the abundance of massive halos and in turn alters the non-linear regime of the density and lensing fields. Previous works studying $\fNLLoc$ have also identified the abundance of massive halos as the key observable difference in the lensing field \citep{Dalal2008ScaleDependentBias, Shirasaki2012fNL, Marian2011fNL, Hilbert2012fNL}. Note, however, that the constraints in Figure \ref{fig:SurveyDep} and Table \ref{tab:FidConstraints} (particularly for LSST) are relatively similar even under scale cuts of $\theta > 10\arcmin$ or $\theta > 20\arcmin$; such cuts are employed in the moments-based weak lensing analyses of \citet{Gatti2020Moments, Gatti2022MomentsDESY3} with DES Y3 data, and mitigate the impact of all lensing-based systematics in DES Y3. These angular scales correspond to physical scales of $10$ to $30 \mpc$ (comoving) depending on the redshift, which are much larger than the virial radius of massive halos and thus correspond to the local, large-scale halo environment. Our constraints after such scale cuts are still comparable/competitive with BOSS/DESI,\footnote{In principle, the scale cuts for LSST may need to be larger than in DES. This is because scale cuts are designed to minimize the impact on the $\chi^2$ of a given datavector. Since the $\chi^2$ increases with improved precision, this implies that for a given systematic with a fixed amplitude, an LSST data-vector will require more scales to be cut compared to a DES data-vector.} and this suggests the analysis' sensitivity to baryon effects is manageable. We discuss this further in Section \ref{sec:ModelChallenge}.

\begin{figure*}
    \centering
    \includegraphics[width = \columnwidth]{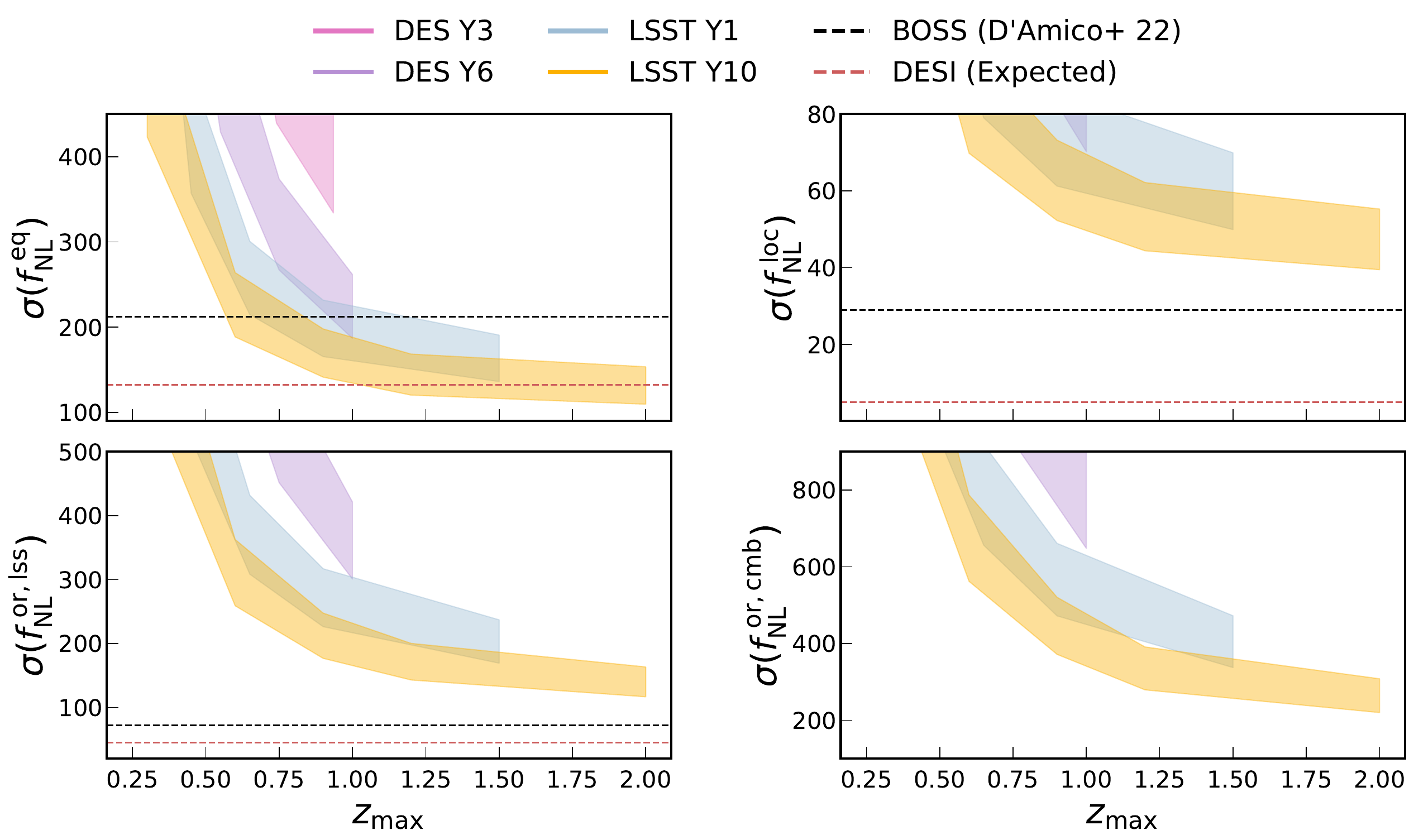}
    \caption{The Fisher information in the 2nd and 3rd moments of the lensing field, shown as a function of the maximum redshift of the data-vector, for four different survey configurations. The maximum redshift is enforced by removing all tomographic bins with a mean redshift of $z_{\rm mean} > z_{\rm max}$. The existing constraints from BOSS are shown as the black line and the potential constraint from DESI are in red. The Fisher information saturates near $z_{\rm max} = 1.25$, due to opposing effects of the lensing amplitude growing towards high redshift and non-linear evolution growing towards low redshift. Note that we only vary the $\fNL$ parameter and do not marginalize over cosmology and IA parameters. This choice is consistent with the BOSS/DESI estimates shown above, which fix the cosmological parameters.}
    \label{fig:RedshiftDep}
\end{figure*}

\textbf{Redshift dependence.} Figure \ref{fig:RedshiftDep} presents the Fisher information for the different surveys, but now limits the maximum redshift of the bins used in the analysis. In each estimate, we use all tomographic bins with a mean redshift below $z_{\rm max}$. Any cross-correlations between bins below $z_{\rm max}$ with those above $z_{\rm max}$ are also not considered. Given the signatures in the lensing field arise from modifications to the halo mass function, the strongest signatures would imprint at low redshift where the structure is most non-linear. However, the weak lensing amplitude depends directly on the amount of matter that lenses the source galaxies. Therefore, high redshift source galaxies contain a larger lensing signal since they are lensed by more foreground structure. These two opposing effects --- the sensitivity of the low-redshift non-linear density field to $\fNL$ but the increased amplitude of weak lensing signals for high-redshift observations --- result in the sensitivity of $\fNL$ from weak lensing saturating at a redshift of $z_{\rm max} \approx 1.25$. This behavior is seen clearly in Figure \ref{fig:RedshiftDep} for LSST Y1 and Y10, while for DES we see similar qualitative trends but do not have tomographic bins with average redshifts beyond $z_{\rm max} \approx 1.25$ to explicitly confirm this. Note, however, that this discussion above concerns signatures (and thus constraints) of \textit{only} $\fNL$. We have verified that in analyses that also marginalize over $\Omega_{\rm m}$, $\sigma_8$, and the IA parameters, the higher redshift bins still add information (though, the choice $z_{\rm max} \approx 1.25$ still approximately maximizes constraints). This improvement is because the high redshift bins, by virtue of their larger lensing signal, have significant information on cosmology and IA parameters and thus improve the \textit{marginalized} constraints on $\fNL$.

\begin{figure}
    \centering
    \includegraphics[width = 0.5\columnwidth]{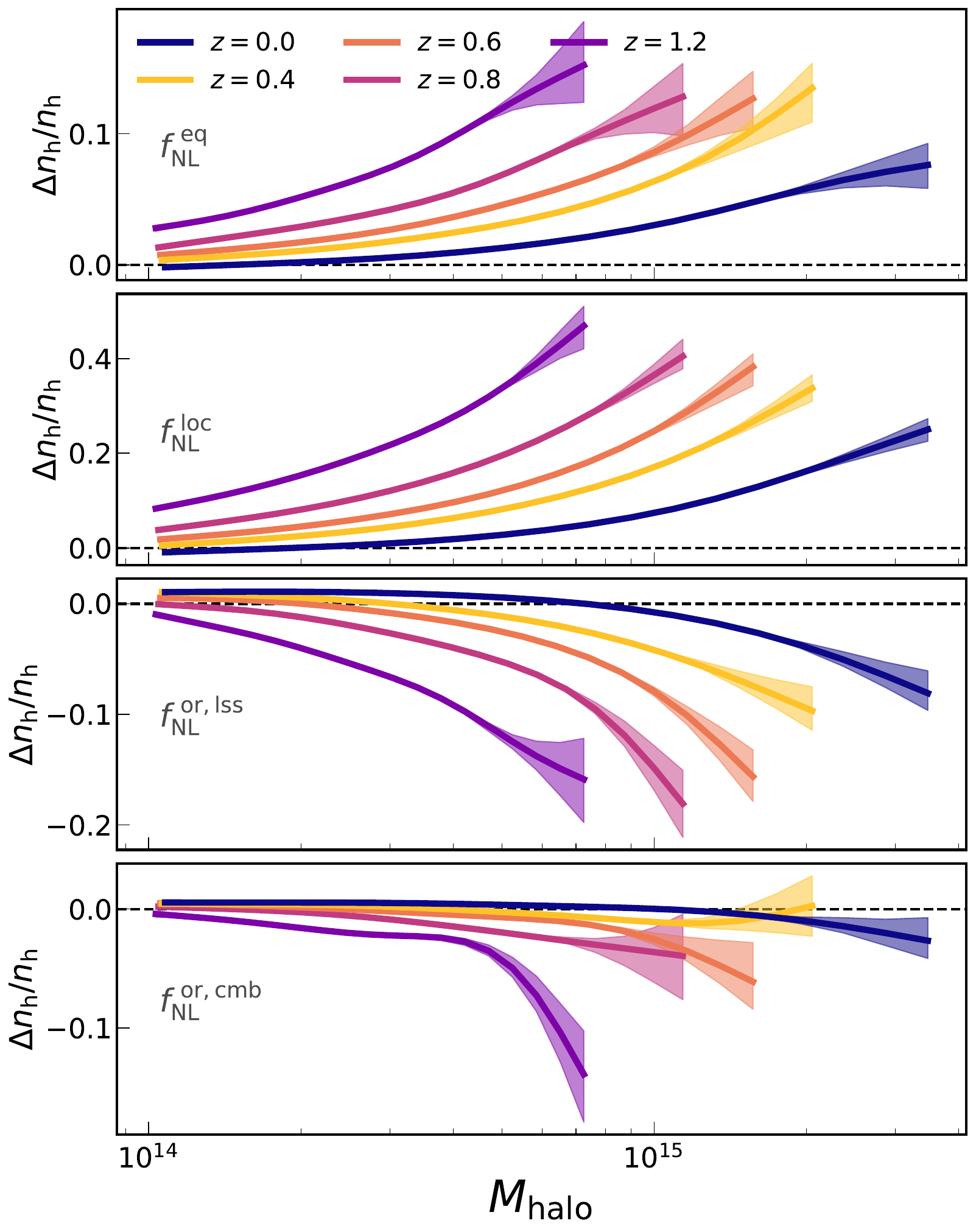}
    \caption{The fractional differences in the halo mass function (HMF) --- which is the counts of halos as a function of mass --- as we vary $\fNL$. The colors show results from different redshifts, and the panels show different $\fNL$ types. The variation is taken as $\ln{\rm HMF}(\fNL = 100) - \ln{\rm HMF}(\fNL = -100)$, which gives us the fractional deviation. The cosmic variance term is suppressed as the simulations with $\fNL = \pm 100$ use the same random seed for the initial conditions.}
    \label{fig:HMF_fNL}
\end{figure}

\begin{figure}
    \centering
    \includegraphics[width = 0.5\columnwidth]{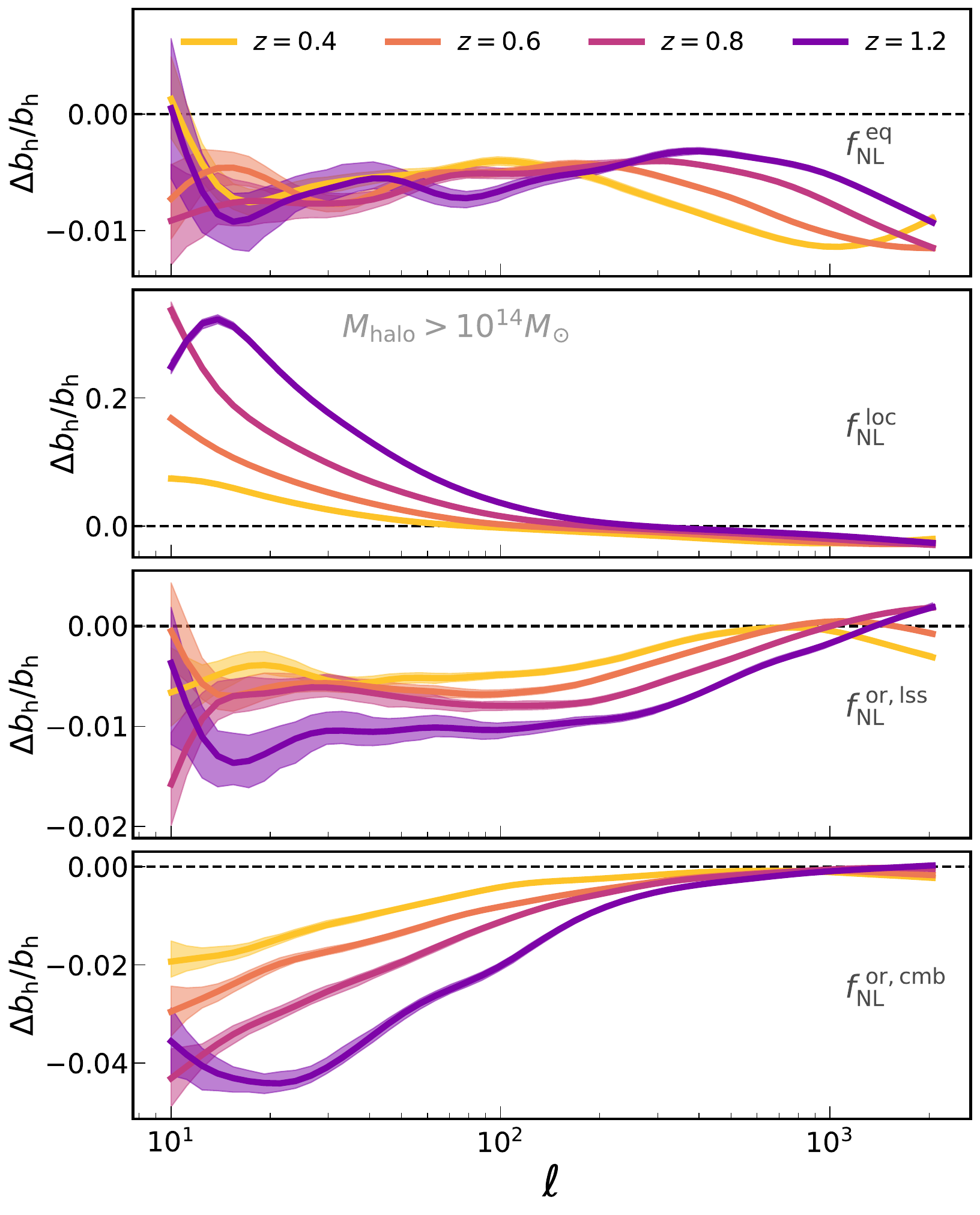}
    \caption{The fractional change in halo bias due to $\fNL$, shown as a function of multipole, $\ell$, for the halo samples of different redshift bins. The variation is computed as $\ln b_h(\fNL = 100) - \ln b_h(\fNL = -100)$. The sample has a fixed lower mass cut of $M > 10^{14}\msol$ at every redshift, and this is set by the resolution limit of the simulation. The bias depends on $\fNL$ over a wide range of scales, including both linear \textit{and} non-linear scales. The cosmic variance term is suppressed as the simulations with $\fNL = \pm 100$ use the same random seed for the initial conditions.}
    \label{fig:Halobias_fNL}
\end{figure}

\textbf{Halo mass function.} We can then also directly explore the variation in the halo mass function (HMF) as we change $\fNL$, as this will then imprint on the weak lensing field. Figure \ref{fig:HMF_fNL} shows the fractional change in the HMF for $\Delta \fNL = 200$, for all four types of $\fNL$, and for five different redshifts. Given the particle resolution of the \textsc{Ulagam} suite and the requirement of at least 100 particles per halo, our halo catalogs are limited to $M > 10^{14} \msol$. While the weak lensing signal is also sensitive to masses below this scale, studying the HMF for such masses is still informative in understanding the qualitative impact of $\fNL$. Figure \ref{fig:HMF_fNL} presents a clear impact of $\fNL$ on halo counts towards the massive halo end. The sign of the fractional change is positive for $\fNLEQ$ and $\fNLLoc$, meaning the abundance of massive halos increases with $\fNL$, and this is expected as for positive $\fNL$ the initial conditions have a larger skewness and have more high-density peaks. At $z = 0$, the HMF for lower mass halos, $M \approx 10^{14} \msol$, is \textit{reduced} with increasing $\fNLLoc$ and $\fNLEQ$. This sign flip is seen more prominently in both orthogonal-type $\fNL$, though the relation is reversed as high $\fNL$ implies a lower halo count. For $\fNLORCMB$, the change is factors of 2-3 smaller than for all the other types. For both orthogonal-type $\fNL$ we once again find that the sign of the change is flipped at redshift $z = 0$, for halos with $M \approx 10^{14} \msol$, as their abundance \textit{increases} with $\fNL$ now. The redshift- and mass-dependence of the sign flip are consistent with findings in \citet{Jung2023fNLHMFQuijote}. This has also been analytically derived in \citet[][see their Section 4.2]{LoVerde2008fNL} and is due to the fact that for a fixed matter energy density, increasing $\fNL$ causes more massive halos to form at the expense of less matter available to form lower mass systems. \citet[][see their Figure 1]{Coulton2022QuijotePNG} also show that the increasing $\fNLOR$ reduces the amplitude of the power spectra on small scales, while increasing $\fNLORCMB$ has a much more mild effect that is nearly non-existent at $z = 0$. While these results are for the power spectrum, they can be translated to signatures in the HMF given halo formation defines the structure of the power spectrum on small scales. Thus, more/less massive halos imply more/less power on small scales.

\textbf{Halo bias.} Given the above discussion on the HMF, a natural conclusion is that the optimal statistic for inflationary constraints from non-linear structure is the HMF itself, where the latter can be measured through the counts of galaxy clusters \citep[\eg][]{DES2020Cluster, Costanzi2021DESxSPT}. While practical considerations motivate weak lensing, in comparison to the HMF, as a simpler observable to robustly analyze, there are theoretical motivations as well. The signature from inflationary models associated with these $\fNL$ types is not solely in the number of high-density peaks of the initial conditions, but also in the way the peaks are spatially clustered. One of the first, well-known signatures of $\fNL$ on the halo field is the scale-dependent bias found in the local-type $\fNL$ \citep{Dalal2008ScaleDependentBias}. This bias goes as $b \propto \fNLLoc/k^2$ and is a diverging signal for $k \rightarrow 0$. Other types of $\fNL$ can also have scale-dependent biases \citep{Schmidt2010BiasPNG}. This bias can imprint on the small-scale density/lensing field in the ``two-halo'' term/regime, which is comprised of the signal from neighboring halos and therefore depends on the halo clustering. We can compute the scale-dependent halo bias in the \textsc{Ulagam} suite as the ratio of the measured halo-matter angular power spectrum and the matter-matter angular power spectrum,
\begin{equation}\label{eqn:halo_bias}
    b_\ell = C_\ell^{hm}/C_\ell^{mm}\,,
\end{equation}
where $C_{\ell}$ is the power at different multipole $\ell$. The choice of using $C_\ell^{hm}$ over $C_\ell^{hh}$ removes the impact of shot noise in our estimate of $b_\ell$. 

Figure \ref{fig:Halobias_fNL} presents the fractional change in the halo bias as a function of $\ell$, for four redshifts and for the four types of $\fNL$. Since this quantity is estimated on mock full-sky maps, and not 3D real-space fields, we do not compute/show the $z = 0$ trends. We reproduce the result of \citet{Dalal2008ScaleDependentBias} for $\fNLLoc$, where the bias of the halo sample grows at large scales (low $\ell$). For $\fNLLoc$ we also see a change in the sign of the bias on small scales to negative values. Similar but weaker features are seen in $\fNLORCMB$, which is expected to have a scaling of $b \propto 1/k$ on linear scales \citep{Schmidt2010BiasPNG}.\footnote{The $\fNLORCMB$ template is an approximation of the full EFT template \citep{Senatore2010WMAP5pngs}. The $b \propto 1/k$ scaling of the former template is considered an artifact of this approximation as the scaling is not found in the latter. In our work, we use $\fNLORCMB$ only to broaden the range of non-Gaussian templates whose signatures we study with weak lensing (and we do not use it to constrain any EFT parameters), in which case differences between the approximate, $\fNLORCMB$ template and the true, EFT template are inconsequential.} Finally, the equilateral type shows the halo bias effect is nearly scale independent for $\ell < 200$; meaning, $\fNLEQ$ simply alters the linear bias. For $\ell > 200$, there is a mild scale-dependent, redshift-dependent behavior, though the amplitude of this variation is still low. The bias for $\fNLOR$ shows similar features to that of $\fNLEQ$,  with a scale-independent bias below $\ell < 200$, and a mildly dependent one above it. The behaviors of the bias on large scales, for all $\fNL$ types, are consistent with the perturbation theory predictions of \citet{Schmidt2010BiasPNG}. 

All scale-dependent signatures of the different $\fNL$ are completely unused if we choose the halo counts as our statistic. On the other hand, the weak lensing field (or any large-scale structure field in general) is sensitive both to the number of massive halos \textit{and} to the halos' spatial clustering properties. Thus, the lensing field utilizes more available information than the halo counts alone. This is also consistent with the lensing field's sensitivity to $\fNL$ even on scales of $\theta > 10\arcmin$, where this angular scale corresponds to $5$ to $20\,\mpc$ across different redshifts, as such scales probe the ``two-halo'' term and the signatures of $\fNL$ imprinted into this term.

Figure \ref{fig:Halobias_fNL} also shows that the impact of $\fNL$ on the halo bias grows with redshift. This, however, is an artifact of selection effects induced by a common halo mass cut across all redshifts. A halo of $M > 10^{14}\msol$ at $z = 2$ is a significantly rarer structure than a halo of $M > 10^{14}\msol$ at $z = 1$. Thus, our fixed mass cut is selecting rarer structures at higher redshift, and therefore the halo bias of the high redshift samples will be larger.

While it may appear contradictory that we discuss the halo bias in a weak lensing field --- which we describe above as an unbiased, direct tracer of the density field\footnote{In practice, intrinsic alignments (see Section \ref{sec:MockMaps}) serve as an analogous ``bias'' term for weak lensing measurements, though some results \citep[\eg][see their Table III]{Secco2022Shear} suggest their impact on the measurements is not yet at a significant level.} --- this is still consistent with our previous statement that the lensing field is insensitive to the galaxy-halo connection. On small scales the clustering of matter can be represented as the combination of three halo properties: their spatial clustering, their number density, and their density profiles. This is denoted the ``halo model'' approach \citep{Cooray2002HaloModel} and utilizes (among other quantities) the halo bias. The galaxy bias and the galaxy-halo connection do not appear in the halo model prediction for the density/lensing field. We decompose the signatures of $\fNL$ in the weak lensing field into signatures in the halo mass function and the halo bias, as this can be a more intuitive picture for understanding the physical origin and scale-dependence of the weak lensing signatures. For example, the scale-dependent bias in the halo field is equivalent to a squeezed bispectrum in the density/lensing field. Both the HMF and the halo bias features discussed above are statistically significant even if we account for cosmic variance, where the latter is generally factors of 3 to 5 larger than the uncertainties shown in Figure \ref{fig:HMF_fNL} and \ref{fig:Halobias_fNL}.

\section{Discussion}\label{sec:Discussion}

Having shown that weak lensing can provide usable constraints on $\fNL$, we now discuss in \S\ref{sec:simlensing} its unique advantage as a probe, and in \S\ref{sec:ModelChallenge} the potential modeling challenges, including existing methods to alleviate them.

\subsection{Advantages of weak lensing as a probe of $\fNL$}\label{sec:simlensing}

Weak lensing has a number of advantages as a cosmological probe, which have aided its development as a leading probe for constraining cosmological parameters \citep[\eg][]{Asgari2021Kids1000, DES2022Y3, More2023HSCY3}. Here, we highlight some of these advantages that are specific to the $\fNL$ signatures we study.

\textbf{Unbiased, direct tracer of the density field.} As mentioned before, the primary advantage of weak lensing as a cosmological probe is that it is a direct tracer of the density field. A vast majority of cosmological signals imprint directly into the correlations of the density field. Historically, the spatial correlations of galaxy fields are a more popular probe as the measurement signal-to-noise is high. However, such analysis requires either marginalization, or prior knowledge, of the galaxy bias parameters. These parameters are required to translate the correlations of the galaxy field, which is the key observable, into those of the density field, which contains the physical signatures of interest. In many cosmological analyses using both lensing and galaxies, the former provides a significant fraction of total constraining power --- even though it is a lower signal-to-noise measurement than the latter; the difference in DES Y3 is 50\% \citep{Monroy2022DESLSS, Secco2022Shear, Amon2022Shear} --- as it does not need to marginalize over the unknown galaxy bias. It is therefore justified to assume weak lensing provides substantial/comparable information on $\fNL$ when compared to the information from galaxy correlations. The results of Section \ref{sec:SurveyDep} confirm this to be the case.

\textbf{Ease of simulation-based modeling.} Any simulations used to model the weak lensing field are defined by two scales: the volume of the survey which sets the simulation size, and the smallest scale in the analysis of the lensing field which sets the simulation resolution. For example, if the analysis does not use scales below $10\mpc$, the simulation can simply be run to be accurate only above those scales. This enables the production of large suites of simulations with coarse resolution that can still be used to infer cosmological constraints. For simulation-based modeling of galaxy fields, however, there is an additional scale in the size of halos/galaxies. Modeling the galaxy field starts from the accurate simulating of \textit{halos}, proceeded by the assignment of galaxies to these halos (using some form of the galaxy--halo connection). Thus, the smallest scale that must be resolved in the simulation is a few times smaller than the smallest halo size. The lack of such a limitation for weak lensing has led to multiple large suites of simulations, some with  $N_{\rm sim} \sim \mathcal{O}(10^3 - 10^4)$, being developed and used in weak lensing analysis \citep{Zurcher2021WLForecast, Zurcher2022WLPeaks, Gatti2023SC, Kacprzak2023Cosmogrid}. This then directly enables the use of non-linear scales in the lensing measurements.\footnote{If the chosen non-linear scales are sufficiently small, then the weak lensing fields will enforce the same small-scale requirement as the galaxy field for simulation-based modeling.} These non-linear scales can currently be modelled only through simulations as analytic approaches are inaccurate in this regime. There are, however, ongoing efforts for \textit{hybrid} approaches that combine the ideas of EFTofLSS with the non-linear predictions of N-body simulations and can thereby extend the range of usable scales \citep{Modi2020HybridEFT, Kokron2022HybridEFT, Banerjee2022HFT}.

\textbf{Constraining galaxy bias parameters with cross-correlations.} The constraints in Figure \ref{fig:SurveyDep} motivate the use of weak lensing-only data as a probe of $\fNL$. However, following the usage of weak lensing in large surveys, we can infer that of equal importance --- if not more --- is the usage of the lensing-galaxy cross-correlation. This correlation constrains, or ``self-calibrates'', the galaxy bias parameters and thus enables/improves the constraints from the galaxy clustering measurements. The latest analyses of $\fNL$ from spectroscopic galaxy surveys utilize a one-loop bispectrum model, which has $\mathcal{O}(10)$ bias parameters compared to the single linear galaxy bias parameter used in most analyses of photometric surveys. Thus, the potential improvement in constraints, due to the self-calibration of bias parameters via the lensing-galaxy cross-correlation, will be greater in this scenario. \citet[][see their Section 7]{Philcox2022BossPNG} also identified that the uncertainty in these bias parameters as the biggest limitation in improving the theoretical modeling of the perturbation theory approach. There are also indications that some assumptions and/or choices of priors for the bias parameters need to be relaxed \citep[\eg][]{Barreeira2022fNLbPhi, Brinch2023EFTPrior}, in which case self-calibration via lensing-galaxy cross-correlations can provide an even larger benefit. Exploring this cross-correlation requires a common modeling framework, and the hybrid-EFT approach mentioned above is a potential method forward.

\textbf{Accessing smaller scales, $\boldsymbol{k > 1 {\rm Mpc}^{-1}}$.} \citet[][see their Figure 4]{Secco2022Shear} show that a lensing measurement at angular scale $\theta$ probes a range of wavenumbers $k$, represented heuristically as
\begin{equation}\label{eqn:ang2k}
    \xi_\kappa(\theta) = \int_0^\infty P_\kappa(k) w(k)\,,
\end{equation}
where $w(k)$ is the weight of each mode, quantifying the sensitivity of measurement $X(\theta)$ to this mode, $P(k)$ is the convergence power spectrum, and $\xi_\kappa(\theta)$ is the convergence 2-point function. We use the same method of \eg \citet{Secco2022Shear}, introduced in \citet{Tegmark2002Theta2k}, to estimate $w(k)$ for the shear 2-point correlation functions --- which corresponds to the 2nd moments measured in our work --- while accounting for the redshift distribution of the different tomographic bins. We will focus on LSST Y10 but note that the results for other surveys are similar. The measurements at the minimum scale of $\theta = 3.2\arcmin$ correspond to mean wavenumbers of $1 < k\,\, [h/{\rm Mpc}] < 2.5$. This is computed as the weighted average of $k$, with weights $w(k)$. If we instead consider the maximum contributing wavenumber --- defined here as the maximum scale with a weight, $w(k)$, that is at least 10\% of the maximum weight $\texttt{max}(w(k))$ --- we find $3.5 < k\,\, [h/{\rm Mpc}] < 6$, depending on the tomographic bin. Even under our conservative angular scale cut of $\theta_{\rm min} = 20\arcmin$, we find the maximum contributing wavenumber is $0.5 < k\,\, [h/{\rm Mpc}] < 2.6$. The CMB analyses of \citet{Planck2020PNGs} probe $\ell_{\rm max} = 2500$, which for the CMB redshift of $z = 1100$ corresponds to scales of $k \lesssim 0.02\,\,h/{\rm Mpc}$. The galaxy correlation function analyses of \citet{Cabass2022MultifieldBOSS, Cabass2022SingleFieldBOSS, Philcox2022BossPNG, Damico2022BossPNG} use up to $k \lesssim 0.2 \,\,h/{\rm Mpc}$, which is set by the current accuracy of the EFT model above this scale. Note that our maximum scale sensitivity will also be limited by modeling choices. However, for angular scale cut of $\theta_{\rm min} = 20 \arcmin$, which is consistent with the choices of DES Y3 \citep{Gatti2022MomentsDESY3}, the model is accurate and the measurement still accesses smaller scales than those of the CMB and galaxy correlations. We discuss in Section \ref{sec:ModelChallenge} the modeling challenges in accessing even smaller scales. The sensitivity of lensing measurements to higher wavenumber than the other probes provides the opportunity to study \textit{scale-dependent} PNGs, especially when combined with other, small-scale measurements from the CMB \citep[][see their Figure 1]{Emami2015ScalDep}. Such scale-dependence is directly connected to inflationary interactions, such as a change in the sound speed $c_s$ over time \citep[\eg][see their Section 2.4.2]{Planck2020PNGs}.

\subsection{Modeling challenges}\label{sec:ModelChallenge}

Simulation-based modeling of weak lensing fields has already been utilized in current surveys to perform precision cosmology \citep{Fluri2019DeepLearningKIDS, Fluri2022wCDMKIDS, Zurcher2022WLPeaks} and many more have used simulations for certain aspects of the modeling pipeline \citep[\eg][]{Secco2022Shear, Amon2022Shear, Gatti2022MomentsDESY3}. However, simulation-based modeling will face new challenges in future surveys, where improved measurement precision requires improved modeling techniques. We detail some of these upcoming challenges below:

\textbf{Higher order shear.} In Section \ref{sec:MockMaps}, we discuss that the observable of a weak lensing survey are the shear fields $\gamma_{1, 2}$. However, this is an approximation as the galaxy shapes actually trace, $e^{\rm obs}_{1,2} = \gamma_{1, 2}/(1 - \kappa)$, where $\kappa$ is the convergence field. In the limit of $\kappa \ll 1$, the expression can be Taylor-expanded to $e^{\rm obs}_{1,2} = \gamma_{1, 2}(1 + \kappa + \kappa^2/2 + \ldots)$. The assumption of $e^{\rm obs}_{1,2} \approx \gamma_{1, 2}$, which we have used in this work, is the ``reduced shear'' approximation. It has been explicitly verified to be a negligible effect for multiple statistics in DES Y3 data \citep{Krause2010ReducedShear, Gatti2020Moments, Anbajagane2023CDFs}. However, simple scaling arguments suggest it will be a statistically significant effect for LSST Y10 precision. Of similar impact is the magnification effect, where more source galaxies are observed in directions with more foreground structure (larger convergence). The impact is modelled as $\propto (1 + q\kappa)$, where $q = \mathcal{O}(1)$. Thus, the magnification impact is similar to that of the reduced shear above, which implies it will also be a notable effect for LSST Y10. Yet another higher-order shear effect is source clustering, which is the correlation of source galaxy positions with the foreground convergence field. This effect arises because source galaxies and lensing convergence are both tracers of the underlying density field, and has been observed in DES Y3 data with different statistics \citep{Gatti2023SC, Anbajagane2023CDFs}. For the 2nd and 3rd moments, its impact can be greatly minimized \citep{Gatti2022MomentsDESY3, Gatti2023SC}. In general, all these effects can be included in (simulation-based) forward modeling approaches at low computational cost.

\textbf{Intrinsic Alignments.} We have already shown the impact of intrinsic alignments in Figure \ref{fig:StatDep}. This test utilized a specific parameterization of IA, with theoretically motivated but fixed parameter values. We do not have observational constraints as the weak lensing data from DES Y3 does not identify an IA signal \citep{Secco2022Shear, Amon2022Shear}. Table \ref{tab:FidConstraints} shows that marginalizing over IA (in addition to $\Omega_{\rm m}$ and $\sigma_8$) leads to LSST Y10-based constraints that are still comparable to DESI for multiple types of $\fNL$. The IA approach we have used, NLA, can be thought of as similar to the hybrid EFT approach, where the underlying framework is that of perturbation theory while the non-linearities in the density field are modelled through N-body simulations. While it is possible to add higher-order terms through the ``Tidal-alignemnt tidal-torquing'' \citep[TATT,][]{Blazek2019TATT} model, the data is currently not precise enough to show a preference for TATT over NLA. The NLA model has been used extensively for various lensing-related analyses \citep[\eg][]{Secco2022Shear, Amon2022Shear, Gatti2022MomentsDESY3}. The requirements for the IA modeling in an LSST Y10 dataset are unclear. If the current, fiducial IA model continues to prove adequate, then we find the IA modeling is not an issue.

\textbf{Other lensing nuisance parameters.} The full analysis of the lensing 2-point correlations also includes marginalizations over other ``nuisance'' parameters that alleviate any systematics-driven biases. These parameters include the mean redshift of the galaxies in each tomographic bin and a multiplicative bias in the estimated shapes of galaxies per bin. The latter is a measurement bias arising primarily from the blending of source galaxies in the image. In current surveys, marginalizing over these parameters leads to negligible impact on the total constraints, in comparison to marginalizing over IA. We can infer this will be more true if we also marginalize over cosmology as we do in this work. It is highly likely that this behavior continues to be the case for LSST, under our current understanding of IA. The effects associated with these nuisance parameters can be easily included in the simulation modeling, and this has already been utilized in multiple analyses \citep[\eg][]{Zurcher2021WLForecast, Fluri2022wCDMKIDS}.

\textbf{Baryon Imprints.} A significant systematic in all analyses related to the density field is the impact of baryonic evolution \citep[\eg][see their Figure 6]{Chisari2018BaryonsPk}. All existing simulation-based models (of either weak lensing or galaxy correlations) employ N-body simulations. While it is possible to use hydrodynamic simulations with galaxy formation models to perform the modeling, such models require many assumptions on the nature of galaxy formation. The assumptions required are often on processes like gas cooling and AGN (Active Galactic Nuclei) feedback, which are the dominant physical processes behind alterations of the density distribution in and around halos \citep{Blumenthal1986AdiabaticContraction, Gnedin2004AdiabaticContraction, Duffy2010BaryonDmProfileDensity, Anbajagane2022Baryons, Shao2022Baryons}. Given the range of possible, allowed assumptions, the simulations manifest a variety of halo property behaviors. Comparative studies show the predictions agree in the overall trends but differ in specific details \citep[\eg][]{Anbajagane2020StellarProp, Lim2021GasProp, Lee2022rSZ, Cui2022GIZMO, Stiskalek2022TNGHorizon, Anbajagane2022Baryons, Anbajagane2022GalaxyVelBias}. Studies on the thermodynamic properties of gas also find differences between the measurements from data and the predictions from these hydrodynamic simulations \citep[\eg][]{Hill2018tSZxGroups, Amodeo2021ACTxBOSS, Pandey2021DESxACT, Anbajagane2022Shocks, Anbajagane2023Shocks}.

An alternative approach to modeling baryons is ``baryonification'' \citep{Schneider2019Baryonification}, which is a flexible, halo-based model that alters the density field in an N-body simulation to include the baryon imprints. This technique provides a higher-level, approximate galaxy formation model that depends only on ``macro'' properties like the halo baryon fraction, the baryon density profiles, dark matter density profile etc. The technique has already been utilized in previous analyses of widefield surveys \citep{Fluri2022wCDMKIDS, Chen2023BaryonY3, Arico2023BaryonY3}. The model flexibility/utility has been shown for the power spectrum/2-point functions --- which are directly related to the 2nd moments we use --- up to $k = 10\,\, h/{\rm Mpc}$ \citep{Schneider2019Baryonification, Giri2021Baryon} and for some higher-order statistics down to $\theta = 1\arcmin$ scales \citep{Lee2023Peaks}, but not for the 3rd moments used in this work. Thus, additional validation work is required to apply this model to the data-vector we employ here. Such baryon correction models will become increasingly necessary for LSST data as simple scale cuts to remove ``baryon contaminated'' measurements, akin to those used in DES Y3, will remove a larger portion of non-linear scales given the increased precision in the LSST data. While Figure \ref{fig:SurveyDep} suggests the LSST Y10 constraints after scale cuts will still be comparable to DESI, these constraints would be much improved by including such non-linear scales and marginalizing over baryon evolution instead.

\section{Conclusion}\label{sec:Conclusion}

We explore the weak lensing field as a potential probe of primordial non-Gaussianities (PNGs), where the amplitude of PNGs is denoted by the $\fNL$ parameter. We consider four types of $\fNL$ --- $\fNLLoc$ which arises from multi-field inflation, $\fNLEQ$ from a strong self-coupling of the inflaton field, $\fNLOR$ and $\fNLORCMB$ from the same physics as $\fNLEQ$ but corresponding to different interactions (see Section \ref{sec:Simulations} for more details) --- and run N-body simulations to extract the Fisher information in DES-like and LSST-like surveys for each type. The \textsc{Ulagam} suite of simulations allows us to forward model wide-field surveys and use physical scales that probe deep into the non-linear regime. Our findings are summarized as follows:

\begin{itemize}
    \item When varying just $\fNL$, the two-point correlation function --- as traced by the 2nd moment of the field --- provides constraints comparable to those from higher-order statistics for $\fNLLoc$ and $\fNLEQ$. However, the latter are clearly better for $\fNLOR$ and $\fNLORCMB$, and for analyses of all $\fNL$ that marginalize over cosmology ($\Omega_{\rm m}$ and $\sigma_8$) and intrinsic alignment parameters (Figure \ref{fig:StatDep}).
    
    \item The shape/configuration information in the lensing field adds significantly to $\fNL$ constraints. The full 3-point correlation function is significantly more constraining than the 3rd moment, which integrates over the shape/configurations (Figure \ref{fig:ConfigDep}). A computationally inexpensive implementation is still challenging.
    
    \item Using the combination of 2nd and 3rd moments as our fiducial statistic, we find the weak lensing-based constraints can be comparable or potentially better than galaxy clustering-based constraints from spectroscopic surveys. For $\fNLEQ$, LSST Y10 (DES Y6) is competitive/better than DESI (BOSS). For $\fNLOR$, the LSST Y10 constraints are a factor of 1.5 to 2 broader than the DESI constraints (Figure \ref{fig:SurveyDep}).
    
    \item The LSST constraints are still comparable to DESI (within factor of 1.5) even with conservative scale cuts of $\theta > 20\arcmin$ (Table \ref{tab:FidConstraints}). At such scales, all systematics associated with weak lensing --- as seen in DES Y3 --- are know to be alleviated.
    
    \item The redshift dependence of the signal peaks for source galaxies of $z \approx 1.25$. Including source galaxies beyond this redshift helps modestly with the constraining power when varying only $\fNL$. However, the high redshift data is valuable in marginalized $\fNL$ constraints, as it helps in constraining the cosmological and IA parameters that are marginalized (Figure \ref{fig:RedshiftDep}).
    \item The impact of $\fNL$ on the lensing field, including the range of scales that probe $\fNL$, can be understood through the impact of $\fNL$ on the halo mass function and the halo bias (Figure \ref{fig:HMF_fNL}, \ref{fig:Halobias_fNL}).
\end{itemize}

Our results show that weak lensing on its own is a useful probe for analyses of $\fNL$, and enhances the search for scale-dependent PNGs. Including cross-correlations between the weak lensing and galaxy fields can result in more significant benefits (see discussion in Section \ref{sec:simlensing}). While a large part of our discussion has centered around Stage IV surveys, such as LSST and DESI, there is also potential for a DES Y6 analysis --- particularly of $\fNLEQ$ (and possibly $\fNLOR$) --- as it is comparable to current constraints from BOSS. This would serve as a pathfinder analysis building towards performing such an analysis with LSST. Improving constraints on $\fNLEQ$ and $\fNLOR$ will help probe the energy scales of inflation and explore one of the energy frontiers in cosmology \citep{Achucarro2022InflationReview}.

We also show that the configuration information in the 3-point correlations is significant. However, the extraction of such information from the traditional weak lensing measurement is not ideal given the weak lensing field is a projected integral of the density field. Methods exist to reconstruct a ``three-dimensional mass map'', where the lensing fields in different tomographic bins are used to reduce the contribution from the foreground/background structure from each bin \citep{Simon2009MassMapp3D, Bernardeau2014ShearNulling}. The final maps are noisier but are better suited for extracting the shape/configuration information, which can improve the $\fNL$ constraints (Figure \ref{fig:ConfigDep}).

Finally, while our work focuses on inflation, and primordial non-Gaussianities in particular, we emphasize that the arguments made above (particularly in Section \ref{sec:Discussion}) can easily extend to any early universe physics that affects the initial conditions. Such physics has thus far been constrained primarily using the galaxy clustering probe, and often with just the two-point function. Weak lensing was not considered a competitive probe of such physics given the limitations in analytic modeling the lensing field, and the reduction in signal amplitude (compared to the amplitude in the 3D density field) due to the projection integral. However, the computing advances of recent times allow for simulation-based modeling to replace the purely analytic approach, and consequently use non-linear scales that are beyond the reach of current analytical modeling approaches. Weak lensing as a probe of new physics must be revisited in light of this shift. In upcoming work, our simulation-based modeling approach will be extended to explore the use of weak lensing in constraining a multitude of different early Universe physics.

\section*{Acknowledgements}

DA is supported by NSF grant No. 2108168. CC is supported by the Henry Luce Foundation and DOE grant DE-SC0021949. HL is supported by the Kavli Institute for Cosmological Physics at the University of Chicago through an endowment from the Kavli Foundation and its founder Fred Kavli.

We thank Will Coulton, Cyrille Doux, Daniel Eisenstein, Giulio Fabbian, Doug Finkbeiner, Josh Frieman, Wayne Hu, Bhuv Jain, Austin Joyce, Nick Kokron, and Lucas Secco for helpful conversations during various stages of this work. We also thank Francisco Villaescusa-Navarro for graciously hosting the \textsc{Ulagam} simulation data and for advice on our public data release. All analysis in this work was enabled greatly by the following software: \textsc{Pandas} \citep{Mckinney2011pandas}, \textsc{NumPy} \citep{vanderWalt2011Numpy}, \textsc{SciPy} \citep{Virtanen2020Scipy}, and \textsc{Matplotlib} \citep{Hunter2007Matplotlib}. We have also used
the Astrophysics Data Service (\href{https://ui.adsabs.harvard.edu/}{ADS}) and \href{https://arxiv.org/}{\texttt{arXiv}} preprint repository extensively during this project and the writing of the paper.

\section*{Data Availability}

The data products of the \textsc{Ulagam} simulations are publicly released and more details can be found at \url{https://ulagam-simulations.readthedocs.io}. These products include the density fields used in this work, alongside the requisite scripts needed to convert them into lensing convergence fields. The halo fields are not yet provided and will instead become public at a later time.

\bibliographystyle{mnras}
\bibliography{References}



\appendix

\section{Validation of simulations} \label{appx:Validation}

We validate here the \textsc{Ulagam} simulation suite through comparisons with theoretical predictions from other models (either based on perturbation theory and/or from simulation-based emulators) and comparisons with the original \textsc{Quijote} suite whose initial conditions are used in the \textsc{Ulagam} runs.

\begin{figure*}
    \centering
    \includegraphics[width = \columnwidth]{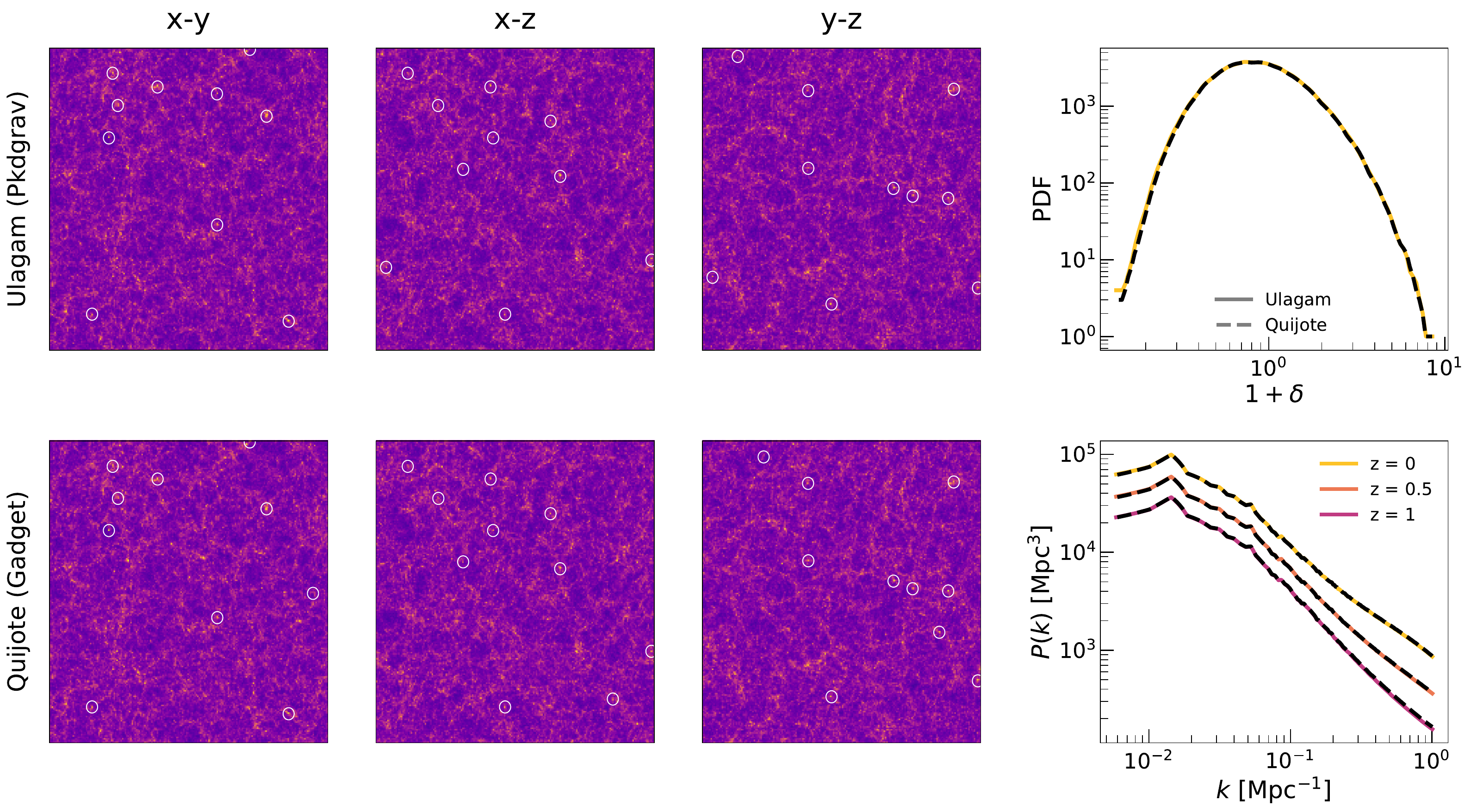}
    \caption{The density fields of a \textsc{Ulagam} and \textsc{Quijote} simulation with the same initial condition. The slices have $500 \mpc/h$ thickness. The density distribution is shown in the top right and the power spectra in the bottom right. White circles show the location of the 10 most massive halos in each slice. There is clear agreement between our simulations and the quijote runs. The specific ordering of the 10 most massive halos varies slightly. The distribution of $1 + \delta$ is within $1\%$ for $\delta > 1$, and grows to $10\%$ for $\delta \approx 0$. The power spectra agree to $1\%$ at z = 0 and grow to $2-3\%$ at higher redshifts, which is within the expected deviations due to differences in N-body solvers \citep{Schneider2016Convergence}}
    \label{fig:QuijoteComparison}
\end{figure*}

\textbf{Comparing \textsc{Ulagam} to \textsc{Quijote}.} Figure \ref{fig:QuijoteComparison} compares the properties of the $z = 0$ three-dimensional density field between a single run in the two simulation suites. While all the \textsc{Ulagam} simulations save lightcone maps and not three-dimensional snapshots, we ran a single simulation that saved the snapshot information as well, specifically to perform this comparison. The left panels of the figure show $500 \mpc/h$ thick slices of the density field projected along different axes (the cartesian axes of the projected field are denoted in the subplot title). The top row shows the \textsc{Ulagam} realization, run with \textsc{Pkdgrav3}, while the bottom row shows the \textsc{Gadget3}-based \textsc{Quijote} realization. Brighter (darker) regions denote overdensities (underdensities). A simple visual comparison shows that the structure in both runs is consistently realized. The white circles denote the location of the top 10 most massive halos in either realization. Most of the locations are common with some minor differences due to the mass ordering of the halos not being completely preserved. This difference in mass-ordering is expected as the halo-finding procedure can be sensitive to stochastic noise. While such effects are suppressed in our comparison given both realizations start from the same initial conditions, differences can still arise due to the use of different gravity solvers with different numerical noise \citep[see][for comparison of power spectra]{Schneider2016Convergence}, and due to different FoF implementations \citep[][see their Figure 5]{Knebe2011HaloFinder}.

The right panels of Figure \ref{fig:QuijoteComparison} shows the probability distribution function of the overdensity field, where the \textsc{Ulagam} and \textsc{Quijote} realizations are within $1\%$ over the vast range of $\delta$ values, and the difference grows to $10\%$ only for $\delta \approx 0$, near the tails of the distribution. The bottom right panel shows the comparison of the power spectra for the two runs at different redshifts. We once again find that the deviations are minimal; they are within $1\%$ at $z = 0$ and grow to $2-3\%$ at higher redshift. Such differences, at our resolution level of $512^3$ particles, are consistent with expectations from comparison studies of the \textsc{Pkdgrav3} and \textsc{Gadget3} \citep{Schneider2016Convergence}.

\begin{figure}
    \centering
    \includegraphics[width = 0.5\columnwidth]{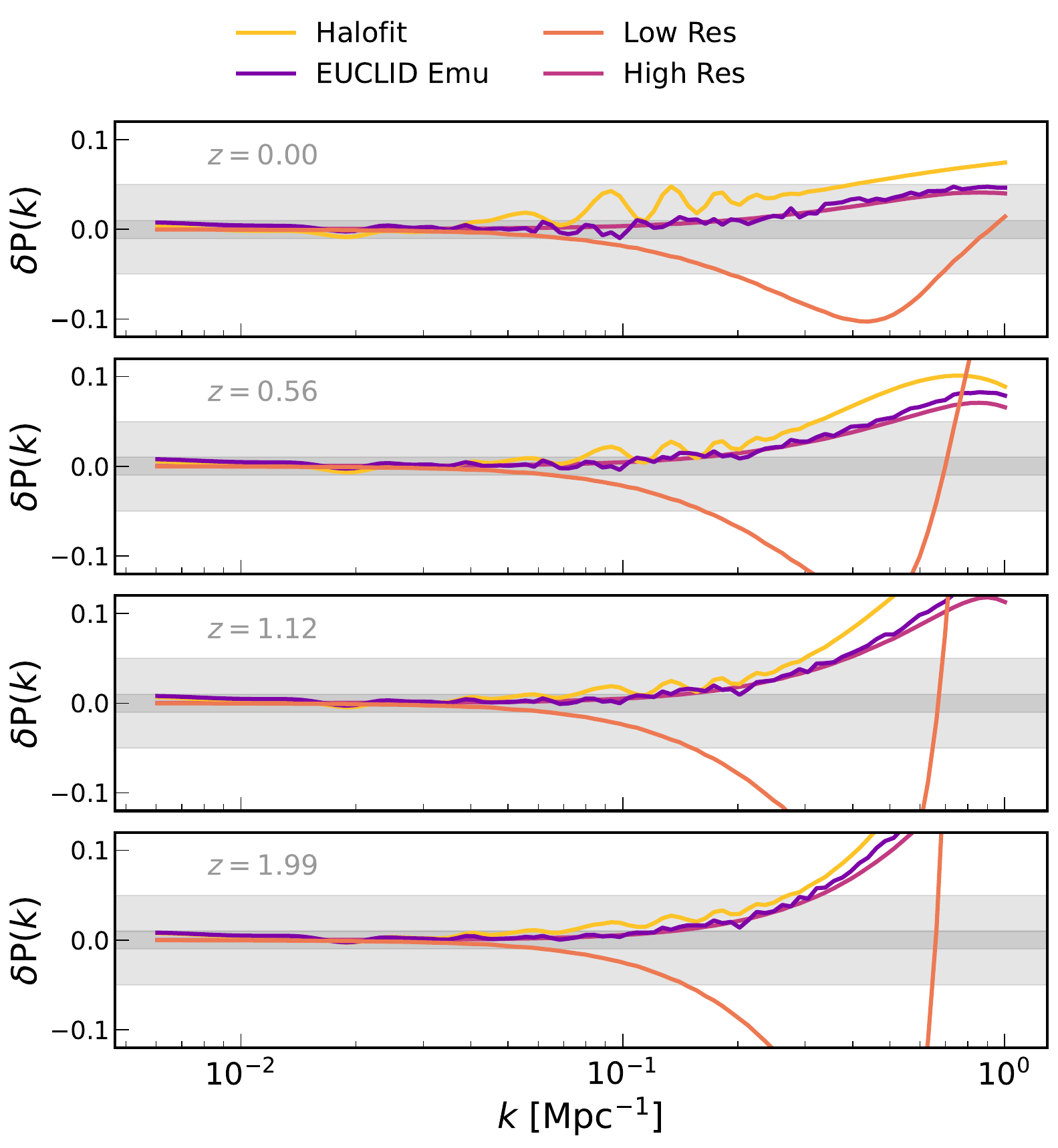}
    \caption{Residuals of matter power spectrum, $\delta P(k) = P_{\rm fid}(k)/P_{\rm truth}(k) - 1$, for five redshifts (shown in top right) where the truth is either computed with \textsc{Class} + \textsc{Halofit}, \textsc{Class} + \textsc{Euclid-Emu} or a lower/higher resolution run with $256^3$/$1024^3$ particles. The fiducial resolution run uses $512^3$ particles. The dark (light) gray band show the 1\% (5\%) residuals. The power spectra match within $5\%$ up to $k = 1$ Mpc$^{-1}$ at z = 0, and $k = 0.3$ Mpc$^{-1}$ at z = 2. Increasing the resolution of the simulation makes the resulting power spectra agree better with both \textsc{Halofit} and \textsc{Euclid-Emu}.}
    \label{fig:Validate_mPk}
\end{figure}

\textbf{Power Spectra.} We then validate the simulations against a number of theoretical models, starting with the power spectrum of the density field. Figure \ref{fig:Validate_mPk} shows the fractional deviation between the \textsc{Ulagam} suite (from the average of five realizations) alongside four different theoretical predictions. Two are simply the average of five lower/higher resolutions runs from the \textsc{Ulagam} suite, while the other two are predictions from \textsc{Halofit} \citep{Takahashi2012Halofit} and \textsc{Euclid-Emu} \citep{Euclid2019Emu}. Both models take the perturbation theory result from the \textsc{Class} boltzmann code \citep{Lesgourgues2011CLASS} and modify it to model the non-linear regime. The agreement with all models (other than the lower resolution run) is within $5\%$ up to $k = 0.3 \mpc^{-1}$ at $z = 2$, and up to $k = 1 \mpc^{-1}$ at $z = 0$. While the deviations with \textsc{Euclid-Emu} increase towards high $k$, improving the simulation resolution to $1024^3$ particles results in significantly better agreement. The \textsc{Euclid-Emu} model was built using \textsc{Pkdgrav3} simulations, but run in a larger volume of $L = 1.9 \gpc$ and with $2048^3$ particles. The \textsc{Halofit} comparison shows oscillatory residuals, particularly at $z = 0$, and this was identified in \citet[][see their Figure 8]{Euclid2019Emu} as a systematic effect in \textsc{Halofit} from not accurately capturing the baryon acoustic oscillation features.

\begin{figure}
    \centering
    \includegraphics[width = 0.5\columnwidth]{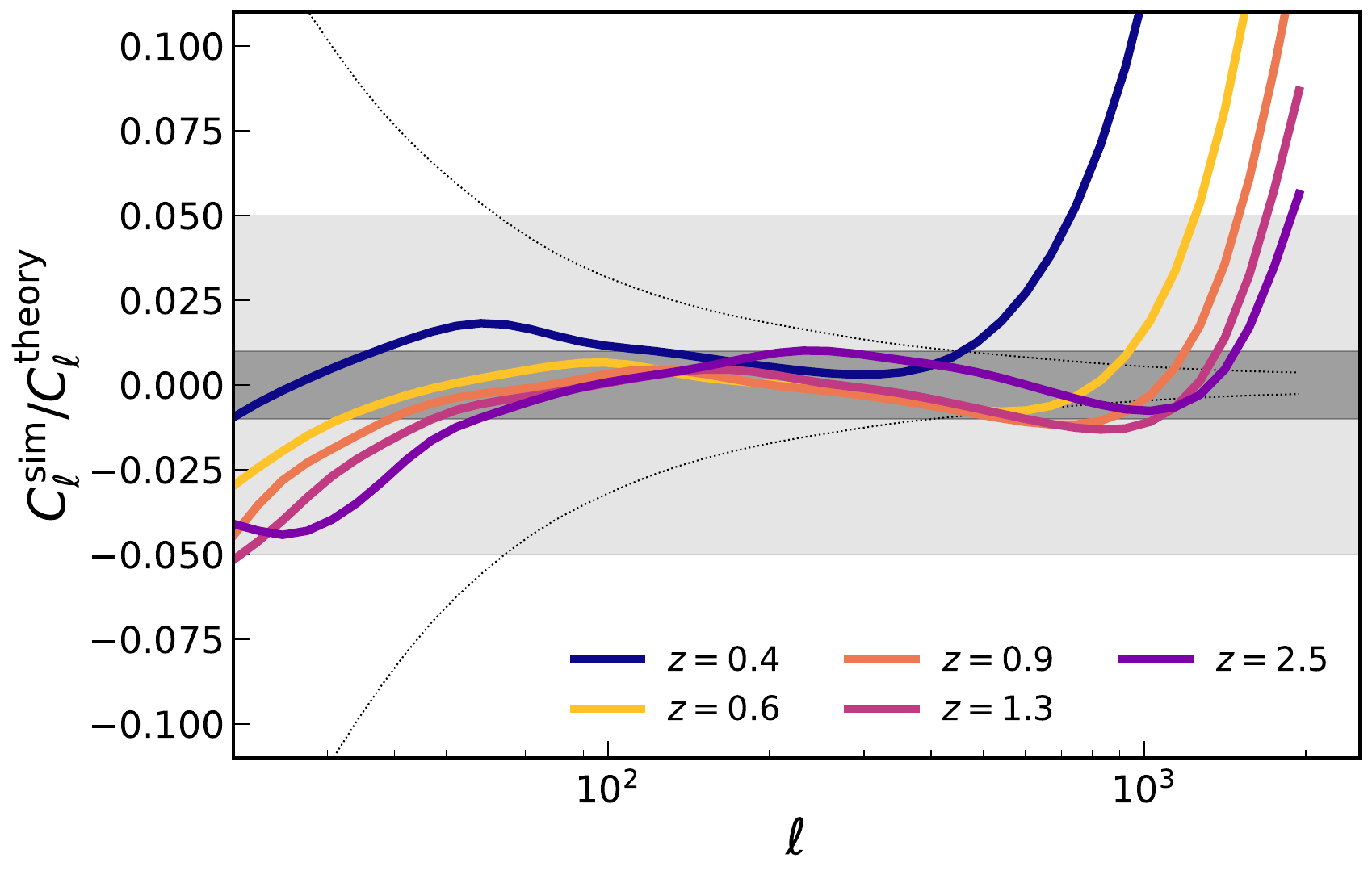}
    \caption{The weak lensing convergence power spectra measured on the full-sky (averaged over all 2000 simulations available at the fiducial cosmology), compared with theoretical predictions from \textsc{Class}, modified by the \textsc{Halofit} prescription for the non-linear regime. The dark (light) gray bands show the 1\% (5\%) error regime. The colored lines show the comparison for different redshifts. The thin black lines show the cosmic variance amplitude at $z = 0.4$. The result for other redshifts is fairly similar, and so we do not show them for brevity. The deviations are within $1\%$ for a wide range of multipoles and redshifts, and grow towards low/high multipoles.}
    \label{fig:Validate_Cl}
\end{figure}

\textbf{Lensing harmonic power spectra.} Figure \ref{fig:Validate_Cl} validates the weak lensing convergence power spectra on the full-sky. We show the comparisons for five source planes at redshifts that span the width of the weak lensing kernel shown in Figure \ref{fig:Nz_Wkernel}. The theoretical model is obtained from \textsc{Class}, modified by the \textsc{Halofit} prescription for the non-linear regime. The dark (light) bands show the 1\% (5\%) error range. The harmonic spectra are the averages of all 2000 full-sky maps, measured by binning the $C_\ell$ in 50 log-space multipole bins between $10 < \ell < 2048$. The deviations are within $1\%$ for a wide range of $\ell$, and increase towards low and high $\ell$.

\begin{figure}
    \centering
    \includegraphics[width = 0.5\columnwidth]{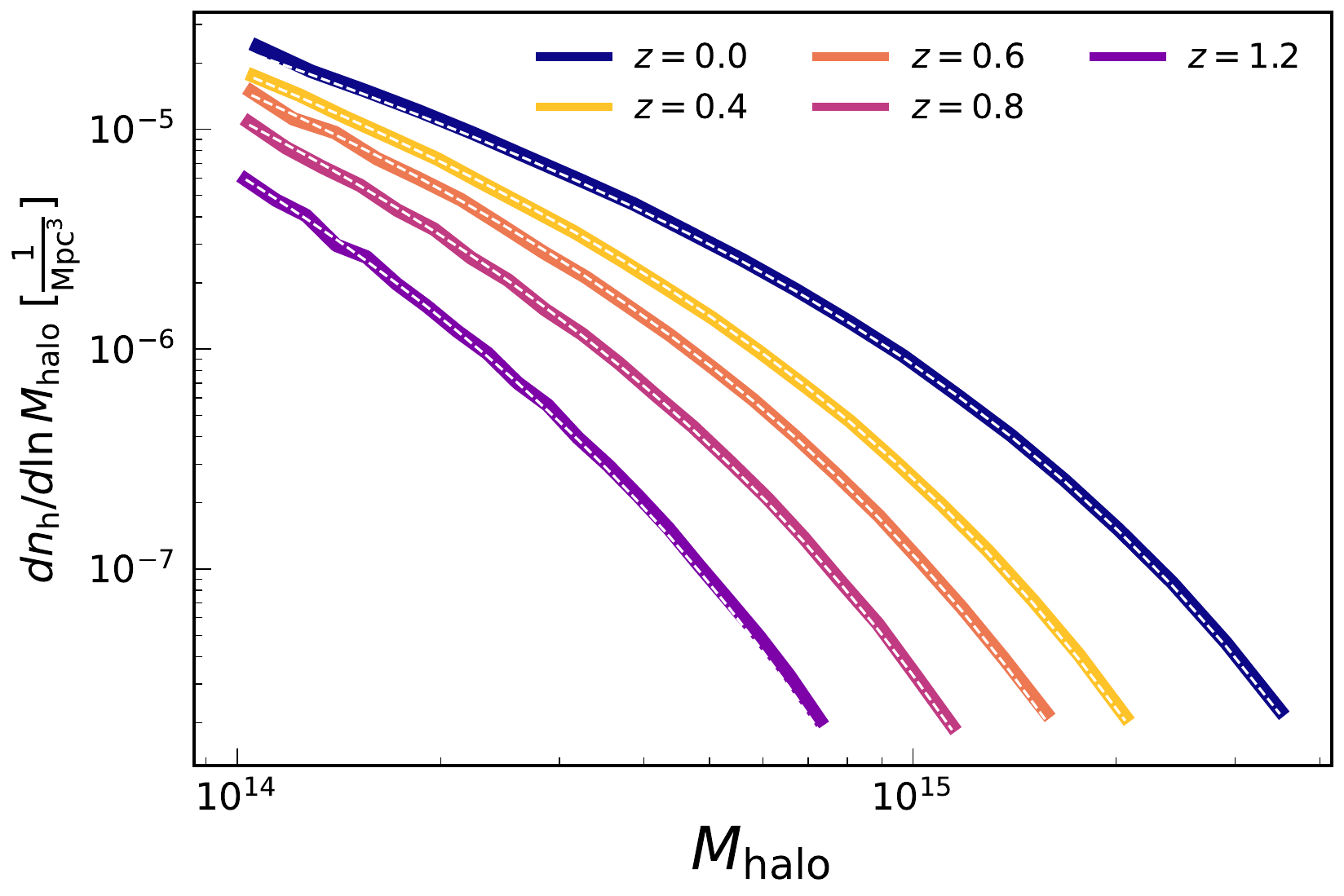}
    \caption{The halo mass function, averaged over 100 realizations of the fiducial cosmology, compared to theoretical predictions from \citet{Watson2013FoFHMF} for halos found through friends-of-friends. The agreement is 5\%  for most redshifts/masses and increases to 10\% for the highest mass objects at each redshift.}
    \label{fig:HMF_validate}
\end{figure}

\textbf{Halo mass function.} We also use the halo abundance in Section \ref{sec:IsolatingInfo} to understand the effect of different $\fNL$. Figure \ref{fig:HMF_validate} shows the HMF averaged over 100 realizations, for different redshifts. The lower mass limit of $10^{14} \msol$ is determined by requiring atleast 100 particles per halo. Overplotted is the theoretical model from \citet{Watson2013FoFHMF} for FoF-based halos, computed using the \textsc{Colossus} package \citep{Diemer2018COLOSSUS}. The agreement is 5\%  for most redshifts/masses and increases to 10\% for the highest mass objects at each redshift.

\begin{figure}
    \centering
    \includegraphics[width = 0.5\columnwidth]{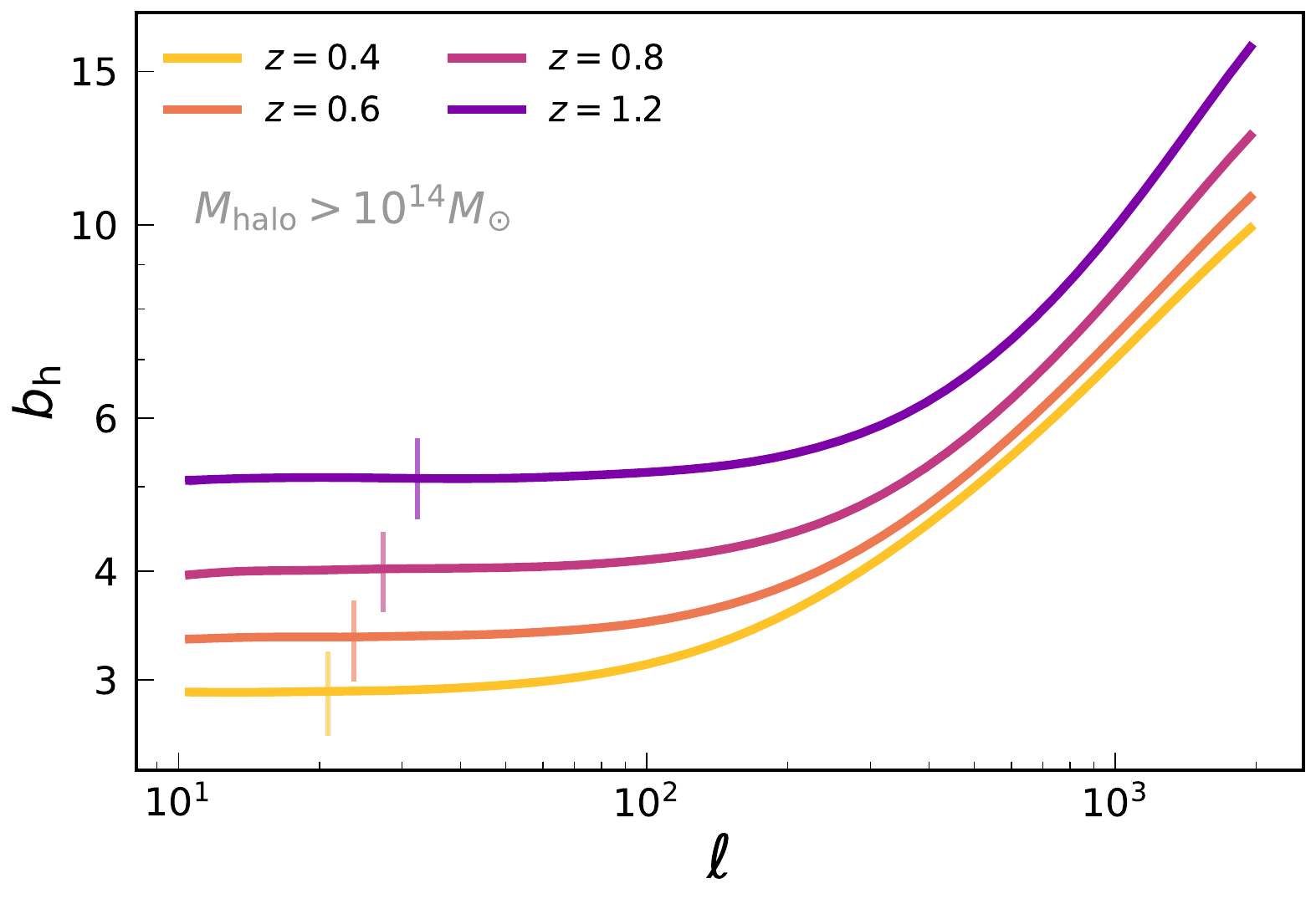}
    \caption{The halo bias, as a function of angular multipole scale and redshift, estimated using Equation \ref{eqn:halo_bias}. We show the average result from 100 realizations at the fiducial cosmology. At large scales (low $\ell$), the bias asymptotes to a constant value, often called the ``linear bias''. The vertical bands show the upper and lower limits from a range of bias predictions \citep{Pillepich2010bias, Tinker2010Bias, Bhattacharya2011Bias, Comparat2017Bias}. The horizontal location of the bands is chosen for visualization purposes and has no other information given the bias is scale-independent on sufficiently large scales. At high $\ell$, there is a clear scale dependence as found in previous work \citep[\eg][]{Mead2021nonlinearbias}.}
    \label{fig:Halobias_validate}
\end{figure}

\textbf{Halo bias.} Finally, we show in Figure \ref{fig:Halobias_validate} the halo bias as a function of multipole, estimated using Equation \ref{eqn:halo_bias}. At large scales, where the bias asymptotes to a constant value, we show the range of predictions from different theoretical models as vertical lines. In specific, we compare the models from \citet{Pillepich2010bias, Tinker2010Bias, Bhattacharya2011Bias, Comparat2017Bias}, and all predictions are computed using \textsc{Colossus} \citep{Diemer2018COLOSSUS}. The exact upper/lower limits used in Figure \ref{fig:Halobias_validate} are from \citet{Comparat2017Bias} and \citet{Tinker2010Bias} respectively. There is a good agreement between the predictions and the measurements across all redshifts. Note that the bias increases with redshift and this is due to the fixed mass cut across all redshifts. For $M > 10^{14} \msol$, the halo sample at $z = 1.2$ is a rarer sample than that at $z = 0$, which then leads to a higher bias. At high $\ell$, the bias grows towards small scales, as has been found in previous work \citep[\eg][]{Mead2021nonlinearbias}.

\section{Numerical convergence of Fisher Information}\label{appx:NumericalConvergence}

\begin{figure*}
    \centering
    \includegraphics[width = \columnwidth]{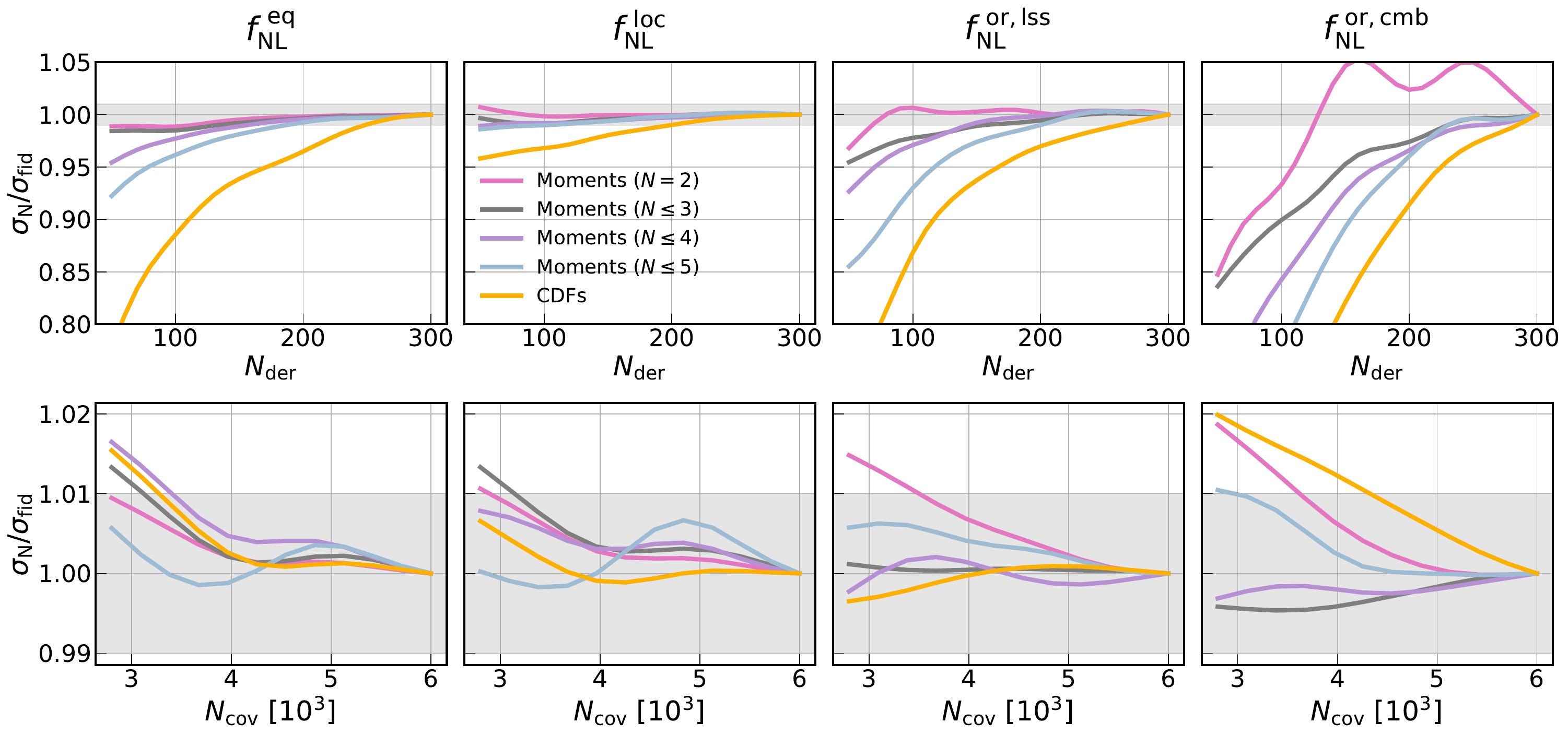}
    \caption{The change in the Fisher information for unmarginalized $\fNL$, for an LSST Y10-like survey, as a function of number of simulations used to estimate the derivative (top) or the covariance (bottom). We show all statistics presented in Figure \ref{fig:StatDep}. The combination of 2nd and 3rd moments --- the fiducial statistic used in this work --- is converged to within 1\%. The convergence properties are similar for the other surveys and for the marginalized $\fNL$ constraints.}
    \label{fig:Convergence}
\end{figure*}

A significant concern in numerical estimates of the Fisher information is the artificial amplification of the information due to numerical noise. This has been documented in previous studies with the \textsc{Quijote} suite \citep[\eg][]{Coulton2022QuijotePNG, Jung2023fNLHMFQuijote}. A simple test of this effect is to vary the number of simulations used in computing either the derivatives or the covariance matrix as used in Equation \ref{eqn:Fisher}. In Figure \ref{fig:Convergence} we present the constraints on $\fNL$, normalized by the fiducial constraints obtained using all available simulations, as a function of the number of realizations used to compute the derivatives and the covariance. We show only the LSST Y10 results, corresponding to the unmarginalized constraints shown in Figure \ref{fig:StatDep}. The constraints for other surveys and for marginalized analyses show the same convergence behaviors as the ones described below. The constraints from the moments (up to 5th order) are well-converged, as an increase in 100 realizations for the derivative calculation results in less than 1\% changes in the constraints. The covariance is also well-converged. Doubling the number of simulations used to estimate the covariance results in less than 2\% differences in the final constraints. Of all $\fNL$ studied here, $\fNLORCMB$ is the most poorly converged, and even then is converged to within 10\% for the combination of 2nd and 3rd moments, which is the fiducial statistic we use in this work.

\section{Full constraints for marginalized $\fNL$}\label{appx:Constraints}

\begin{figure*}
    \centering
    \includegraphics[width = \columnwidth]{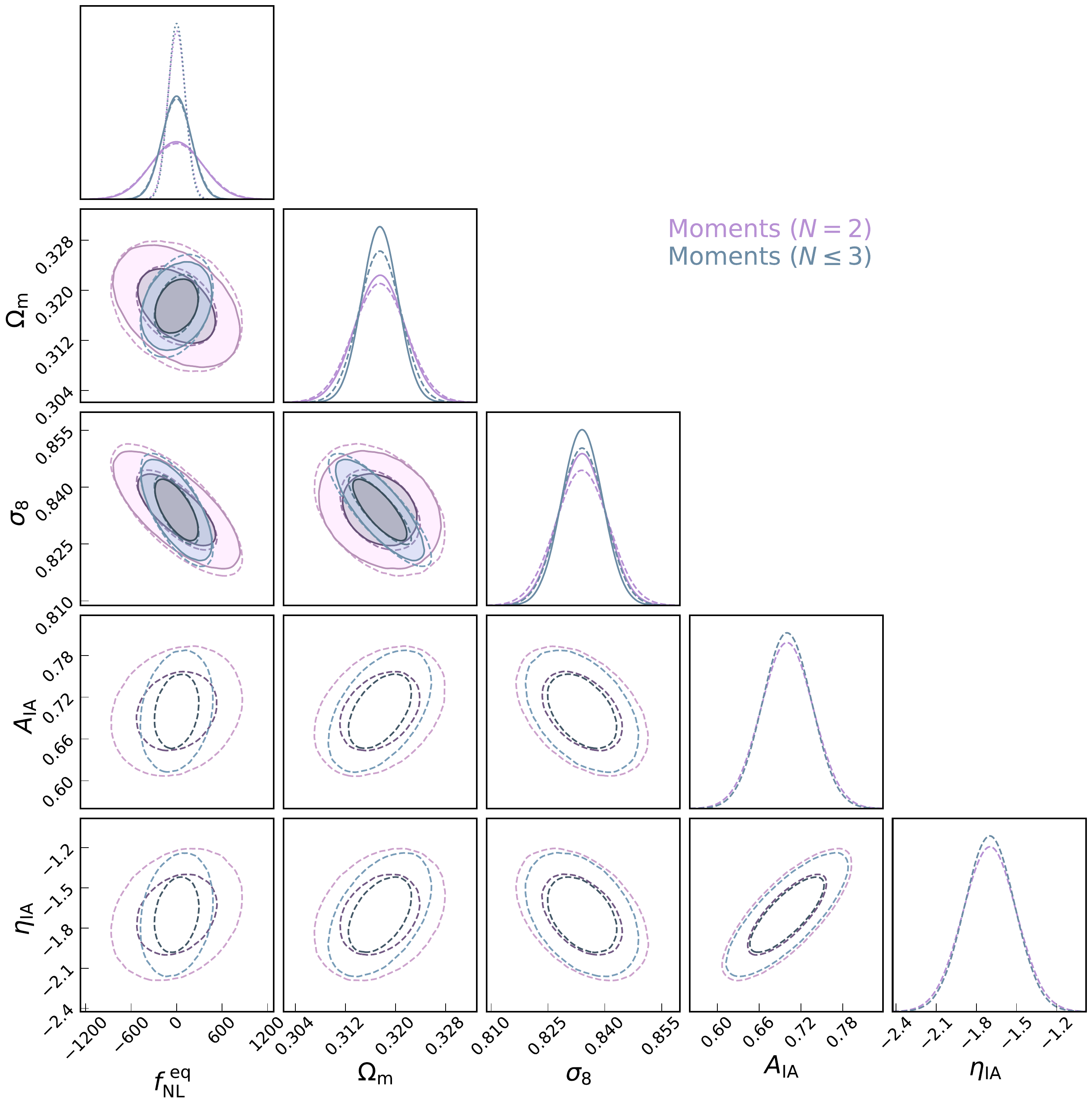}
    \caption{The full Fisher constraints for $\fNLEQ$, $\Omega_{\rm m}$, $\sigma_{8}$, $A_{\rm IA}$ and $\eta_{\rm IA}$. The dashed contours show constraints over all five paramters, while solid contours are constraints varying $\fNL$ and cosmology parameters alone. There is only marginal differences between the two. The 1D posterior for $\fNLEQ$ also shows the unmarginalized constraint in the dotted lines. }
    \label{fig:Triangle_EQ}
\end{figure*}

For completeness, we show the full triangle plot of constraints for the marginalized $\fNLEQ$ analysis. The $\sigma_8 - \fNLEQ$ plane shows an anti-correlation; increasing $\fNLEQ$ leads to more structure on small scales, which can be partially compensated by a reduction in $\sigma_8$ as that removes structures across a wide range of scales. A similar anti-correlation is found in $\Omega_{\rm m} - \fNLEQ$. However, this is only for the analysis of the 2nd moments; the $\Omega_{\rm m} - \fNLEQ$ correlation is positive for analyses combining the 2nd and 3rd moments.

We do not show the full constraints from the other parameters for brevity. Note that in this work we do not vary all $\fNL$ parameters at once, and instead analyze them one at a time; the sole exception is the estimate of the sound speed, $c_s$, in Section \ref{sec:SurveyDep} where we vary both $\fNLEQ$ and $\fNLOR$. The constraints for $\fNLLoc$ have qualitatively similar degeneracies in all parameter planes as the ones described above. The degeneracies of $\fNLOR$ and $\fNLORCMB$ with IA and cosmological parameters are qualitatively different, and this can be inferred from the behavior of the halo mass function in Figure\ref{fig:HMF_fNL}. In particular, the sign of the correlation between $\fNL$ and the cosmology parameters is reversed in every plane. Figure \ref{fig:HMF_fNL} shows that the orthogonal-type $\fNL$ \textit{reduce} the number of high-density peaks --- which is also supported by the analysis of \citet{Coulton2022QuijotePNG}, which found the change in the power spectrum is also negative --- and therefore, the sign of the correlation will be flipped.


\label{lastpage}
\end{document}